\documentclass[fleqn,usenatbib]{mnras}

\usepackage{newtxtext,newtxmath}

\usepackage[T1]{fontenc}
\usepackage{float}
\DeclareRobustCommand{\VAN}[3]{#2}
\let\VANthebibliography\thebibliography
\def\thebibliography{\DeclareRobustCommand{\VAN}[3]{##3}\VANthebibliography}

\usepackage{graphicx}	

\title[Two c's in a pod]{Two c's in a pod: Cosmology independent measurement of the Type Ia supernova colour - luminosity relation with a sibling pair}

\author[R.~Biswas~et~al.]{
Rahul~Biswas,$^{1}$\thanks{E-mail: rbiswas4@gmail.com }
Ariel~Goobar,$^{1}$\thanks{E-mail: ariel@fysik.su.se }
Suhail~Dhawan,$^{1,2}$
Steve~Schulze,$^{1}$
Joel~Johansson,$^{1}$
\newauthor
Eric~C.~Bellm,$^{3}$
Richard~Dekany,$^{4}$
Andrew~J.~Drake,$^{5}$
Dmitry~A.~Duev,$^{5}$
\newauthor
Christoffer~Fremling$^{5}$
Matthew~Graham,$^{5}$
Young-Lo~Kim,$^{6}$ 
Erik~C.~Kool,$^{7}$
\newauthor
Shrinivas~R.~Kulkarni,$^{5}$
Ashish~A.~Mahabal,$^{5,8}$
Daniel~Perley,$^{9}$ 
Mickael~Rigault,$^{6}$
\newauthor
Ben~Rusholme,$^{10}$
Jesper Sollerman,$^{7}$
David~L.~Shupe,$^{10}$
Matthew~Smith,$^{6}$
Richard~S.~Walters$^{4}$
\\
$^{1}${Oskar Klein Centre, Department of Physics, Stockholm University, SE 106 91 Stockholm, Sweden}\\
$^{2}${Kavli Institute for Cosmology, University of Cambridge, Madingley Road, Cambridge CB3 0HA, UK}\\
$^{3}${DIRAC Institute, Department of Astronomy, University of Washington, 3910 15th Avenue NE, Seattle, WA 98195, USA}\\
$^{4}${Caltech Optical Observatories, California Institute of Technology, Pasadena, CA  91125, USA}\\
$^{5}${Division of Physics, Mathematics, and Astronomy, California Institute of Technology, Pasadena, CA 91125, USA}\\
$^{6}${Univ Lyon, Univ Claude Bernard Lyon 1, CNRS, IP2I Lyon / IN2P3, IMR 5822, F-69622, Villeurbanne, France}\\
$^{7}${Department of Astronomy, Oskar Klein Centre, Stockholm University, Albanova, 10691 Stockholm, Sweden}\\
$^{8}${Center for Data Driven Discovery, California Institute of Technology, Pasadena, CA 91125, USA}\\
$^{9}${Astrophysics Research Institute, Liverpool John Moores University, IC2, Liverpool Science Park, 146 Brownlow Hill, Liverpool L3 5RF, UK}\\
$^{10}${IPAC, California Institute of Technology, 1200 E. California Blvd, Pasadena, CA 91125, USA}\\
}
\date{Accepted XXX. Received YYY; in original form ZZZ}

\pubyear{2021}
\usepackage{lineno}
\usepackage{graphicx}
\usepackage{xspace}
\usepackage{color}
\usepackage{units}
\usepackage{hyperref}
\usepackage{float}
\usepackage{multibib}
\usepackage{lipsum}
\usepackage{tabularx}

\newcommand{\snia}{SN Ia}
\newcommand{\sneia}{SNe Ia}
\newcommand{\sn}{SN}
\newcommand{\sne}{SNe}

\providecommand{\arcsec}{\ensuremath{''}\xspace}

\newcommand{\kpc}{\unit{kpc}}

\newcommand{\LCDM}{\ensuremath{\rm \Lambda CDM}\xspace}

\newcommand{\be}{\begin{equation}}
\newcommand{\ee}{\end{equation}}
\newcommand{\beqn}{\begin{eqnarray}}
\newcommand{\eeqn}{\end{eqnarray}}

\begin{document}
\label{firstpage}
\pagerange{\pageref{firstpage}--\pageref{lastpage}}
\maketitle

\newcommand{\oldsn}{{AT 2019lcj}}
\newcommand{\newsn}{{SN 2020aewj}}
\newcommand{\ztfname}{ZTF19\lowercase{aambfxc}}
\newcommand{\zuds}{\texttt{ZUDS}}
\newcommand{\salt}{\texttt{SALT2}}
\newcommand{\wise}{\texttt{WISE}}
\newcommand{\package}{package}
\newcommand\refcor[1]{\textcolor{black}{#1}}
\newcommand\changes[1]{\textcolor{black}{#1}}
\newcommand\revtwo[1]{\textcolor{black}{#1}}
\newcommand\ag[1]{\textcolor{black}{#1}}

\begin{abstract}
	Using Zwicky Transient Facility (ZTF) observations, we identify a pair of "sibling" Type Ia supernovae (\sneia{}), i.e., hosted by the same galaxy at $z=0.0541$. They exploded within 200 days from each other at a separation of $0.6\arcsec\ $corresponding to a projected distance of only \refcor{ $0.6\ \kpc.$}  Performing \salt\ light curve fits to the $gri$ ZTF photometry, we show that for these equally distant `standardizable candles'',\ there is a difference of 2 magnitudes in their rest frame $B$-band peaks, and the fainter \sn\ has a significantly red \salt\  colour $c$ = 0.57 $\pm$  0.04, while  the \refcor{stretch values $x_1$ of the two \sne\ are similar,} suggesting that the fainter \sn\ is attenuated by dust in the interstellar medium of the host galaxy.  We use these measurements to infer the SALT2 colour standardization parameter, $\beta = 3.5 \pm 0.3$, independent of the underlying cosmology and Malmquist bias. Assuming the colour excess is entirely due to dust, the result differs by $2\sigma$ from the average Milky-Way total-to-selective extinction ratio, but is in good agreement with the colour-brightness corrections empirically derived from the most recent \snia\ Hubble-Lemaitre diagram fits. Thus we suggest that SN "siblings", which will increasingly be discovered in  the coming years, can be used to probe the validity of the colour and lightcurve shape corrections using in \snia\ cosmology while avoiding important systematic effects in their inference from global multi-parameter fits to inhomogeneous data-sets, and also help constrain the role of interstellar dust in \snia\ cosmology. 
\end{abstract}


\section{Introduction}
The discovery of the late-time accelerated expansion of the universe using Supernovae of Type Ia (\sneia\ ) as cosmological yardsticks \citep{1998AJ....116.1009R, 1999ApJ...517..565P} had a profound impact on our understanding of the cosmic composition. The pioneering work of the Supernova Cosmology Project and the High-z Supernova Search Team was followed by many efforts to improve the use of SNe Ia in cosmology with the purpose to better understand the nature of dark energy \citep[see][for a review]{2011ARNPS..61..251G}. Essential for the standardization of SNe Ia to obtain precise distances are the corrections for the lightcurve shape-brightness relation \citep{1993ApJ...413L.105P} and the colour-brightness relation \citep{1998A&A...331..815T}. In recent years, most \sneia\ cosmological samples are analyzed in the \salt\ lightcurve framework \citep{2005A&A...443..781G, 2007A&A...466...11G,2010A&A...523A...7G}. The distance modulus $\mu$ is corrected for lightcurve shape ($x_1$) and colour ($c$) as 
$$\mu = m - M + \alpha \cdot x_1 - \beta \cdot c,$$
\refcor{where $\alpha$ and $\beta$ are constants, whose values are determined by fitting to a Hubble-Lemaitre diagram. }
The colour measurement, $c$, which corresponds approximately to $E(B-V)$ where the \salt\ template SED is used as reference is thus multiplied by an empirically derived parameter $\beta$, the topic of this work. In the \salt\ framework, the absolute magnitude $M_B$ in rest-frame $B$-band is used as the  
anchoring point. The \salt\ model performs standardization in a two step process. First the \salt\ model is fitted to the light curve data of each supernova. These parameters and uncertainties are then used to simultaneously determine the parameters $\alpha$ and $\beta$ along with the cosmology. Traditionally, this was performed for each cosmological model (and possibly) with complementary data if desired. This made the standardized distance moduli (through the values of $\alpha\ $ and $\beta$) dependent on both the choice of the cosmological model used, and the complementary data. The use of \texttt{SALT2mu}~\citep{2011ApJ...740...72M}, which uses piece-wise continuous cosmological distance moduli functions of \LCDM\  in different redshift bins, ameliorates the cosmological model dependence. More importantly, it paves the way for a generalized intrinsic scatter model in the form of a covariance \refcor{between the parameters $\{m, x_1, c \}$.} \\
\ag{It may be tempting to view the colour-brightness relation as being entirely due to interstellar extinction by dust in the host galaxy of the supernovae, i.e., with $\beta$ corresponding to the total-to-selective extinction parameter, $R_B$, following the analogy of Milky-Way extinction \citep[see e.g.,][]{1989ApJ...345..245C}. However, 
when $\beta$ is fitted using the ensemble of low and high-$z$ SNe Ia to minimize the scatter in the Hubble-Lemaitre diagram residuals, its value comes out to be significantly lower than $\beta \sim R_B\approx 4.1$, the Milky-Way average value \citep{2006A&A...447...31A,2009ApJS..185...32K,2010ApJ...716..712A,2012ApJ...746...85S,2014A&A...568A..22B,2018ApJ...859..101S}. Focusing on the most recent analyses, \citet{2014A&A...568A..22B} find $\beta = 3.101\pm 0.075$ based on a sample of 740 \snia\, and \citet{2018ApJ...859..101S} find $\beta = 3.030 \pm 0.063$ when they extend the sample with newer discoveries, totalling 1048 SNe with similar lightcurve selections. 
Since only objects with moderate colour have been kept in the samples used for cosmology,  $c \le 0.3$, it has been argued in those studies that the low values of $\beta$ could be mainly due to intrinsic colour variations \citep[see also][]{2011A&A...529L...4C}. }

\ag{An additional complication to the standardization of SNe Ia magnitudes found over the past decade decade is that that}  there is a significant environmental dependence of the distance modulus on the the properties of the host galaxy beyond the colour and lightcurve shape corrections \citep{2010ApJ...715..743K,2010MNRAS.406..782S,2010ApJ...722..566L, 2013ApJ...770..108C,2015ApJ...802...20R,2020A&A...644A.176R,2021MNRAS.501.4861K}. Moreover, it has been realized that selection effects in observational surveys result in incompleteness in the distribution of \sneia\ properties due to interaction with the intrinsic dispersion\refcor{, and affect the inferred values of $\beta.{}$ }In cosmological analyses\refcor{~\citep{2009ApJS..185...32K,2014A&A...568A..22B,2019ApJ...874..150B}} these are usually corrected through \refcor{a set of bias corrections terms~\citep{2014ApJ...793...16M,2017ApJ...836...56K,2019MNRAS.485.1171K,2021arXiv210201776P}} based on simulations. Such simulations require detailed inputs of population models inferred from the data~\citep{2013ApJ...764...48K,2016ApJ...822L..35S} and a detailed description of the observational procedure. Given the complexity of such a program and the importance of these parameters to cosmology, complementary checks which do not involve many of such effects like environmental dependence or population models are important cross-checks.\\
\ag{
Recently, \citet{2020arXiv200410206B} suggested that the $\beta$ is mainly due to dust, but that the extinction properties of \sneia\ depend on the host galaxy stellar mass, thus providing further uncertainty in the reported single "universal" values of $\beta$. \citet{2021arXiv210506236J} reach a similar conclusion \citep[but see][for a different view]{2021arXiv210205678T}.} \\

\ag{In summary, }concerns have been raised that the colour-brightness parameter $\beta$ derived from cosmological analysis may be biased due to selection effects, procedural mistakes, degeneracies with other parameters in the global fits, redshift uncertainties, K-corrections, calibration errors, and possibly even Milky-Way extinction errors. Indeed, over time and for different samples, the reported best fit value of $\beta$ has varied from $1.57\pm 0.15$~\citep{2006A&A...447...31A}, $2.47 \pm 0.06$~\citep{2012ApJ...746...85S}, to $\beta \approx 3.0$, reported by \citet{2018ApJ...859..101S}. 
\ag{Ideas to reconcile the low values of $\beta$ with non-standard extinction have been put forward, e.g.,  that the dimming dust is localized to the circumstellar environment \citep{2005ApJ...635L..33W,2008ApJ...686L.103G}. The latter suggestion 
has been explored studying the wavelength dependent attenuation of SN 2014J, a highly reddened \snia\ in the nearby galaxy M82 which also showed non-standard extinction \citep{2014ApJ...784L..12G,2014ApJ...788L..21A,2014MNRAS.443.2887F}, and through searches of emission from heated circumstellar dust \citep{2015MNRAS.452.3281M,2017MNRAS.466.3442J}. The colour relations of samples of nearby reddened SNe Ia have been reporting "non-standard" extinction laws for over a decade \citep[see e.g,][]{2008A&A...487...19N}, and have recently been expanded with observations ranging from UV to NIR
\citep{2014ApJ...789...32B,2015MNRAS.453.3300A}. The conclusion from these studies is that extinction by circumstellar dust likely plays a minor role in the observed colour-brightness relation of SNe Ia, and that a diverse population of dust in the interstellar medium of other galaxies is required to explain the observations, even after intrinsic colour variations are taken into account. An intriguing possibility that has been put forward is that dust grains may be fragmented by collisions between dust clouds, as these are accelerated by radiation pressure from the SN itself \citep{2017ApJ...836...13H,2018MNRAS.479.3663B}, or by radiation in the interstellar medium \citep{2021ApJ...907...37H}. 
} \\

While the well-measured nearby SNe Ia, too close to be in the Hubble flow,  were able to provide supporting evidence for the "non-standard" wavelength dependence of attenuation, they could not probe the absolute dimming. This work bridges the efforts between the local and cosmological efforts to study the relation between colour excess and brightness attenuation for SNe Ia. 
\ag{Our measurement of $\beta$ based on siblings uses the same SALT2 tools as the cosmological analysis, but is free from the potential systematic effects that are often attributed to the Hubble-diagram inferences of $\beta$. We emphasize that our analysis is completely agnostic as to the origin of the colour-brightness relation.}

\ag{
\section{ZTF and the case for SNIa Siblings }
}
As time domain astronomy surveys get larger and run for longer survey duration, they can find rare objects. Multiple \sneia\  occurring in the same host galaxy, known as siblings. Recent surveys have increased the number of such siblings, allowing ensemble level questions to be addressed through them~\citep{2020ApJ...896L..13S,2020ApJ...895..118B}, which use the Dark Energy Survey and CSP data to discuss to what extent sibling supernovae share common properties. 

\ag{
The Zwicky Transient Facility (ZTF) survey \citep{2019PASP..131a8002B, 2019PASP..131g8001G} is a $3\pi$ imaging survey of the Northern sky conducted on the (48-inch) Samuel Oschin Telescope at the Palomar Observatory \citep[see][for a more detailed description of the survey specifications]{2019PASP..131f8003B} between 2018 and 2020, later replaced by ZTF-II. It included a public survey in $g$ and $r$ bands with a nominal $5-\sigma$ depth of $\sim 20.5$ mag with a 3-day cadence, along with a few programs run by the ZTF partnership including an extragalactic survey in the $i$-band with 4-day cadence, designed to obtain a three-filter lightcurve sample of Type Ia supernovae for cosmological applications. \refcor{The procedure for processing the survey data and the data products are described in~\citet{2019PASP..131a8003M}}. As part of the public survey, the Bright Transient  Survey (BTS)~\citep{2020ApJ...895...32F} is an effort aimed at collecting an untargeted, nearly complete, magnitude limited sample of spectroscopically classified transients reaching 18.5 mag.}\\

As ZTF scans the sky with unprecedented speed and depth \ag{it has the potential to discover sibling supernovae, as discussed in Graham et al., 2021, (in prep) \citep[see also][for another reported candidate sibling pair]{2021arXiv210309937S}}. In this work, we use a particular set of sibling \sne\  Ia, found at nearly identical positions within a year, but with significantly different observed flux and colours \ag{to constrain the parameter $\beta$ in a cosmology independent way, and yet without any additional assumptions on the origin of the colour excess}. Our precision based on this single system is only slightly weaker than constraints obtained analysing $\sim 10^3$ SNe in the Hubble-Lemaitre diagram. 
This can be improved with a number of sibling supernovae, and with different properties of the siblings, can also provide a cosmology independent constraint on both $\beta$ and the lightcurve width-brightness correction factor, $\alpha$. Thus, it is a complementary source of information for supernova cosmology, and can be also used to probe extinction in these systems even if a wide lever arm in wavelength range is missing.\\

\section{\ztfname\ : the tale of two supernovae}
\label{sec:data}
\begin{figure}
    \begin{center}
    \includegraphics[width=0.45\textwidth]{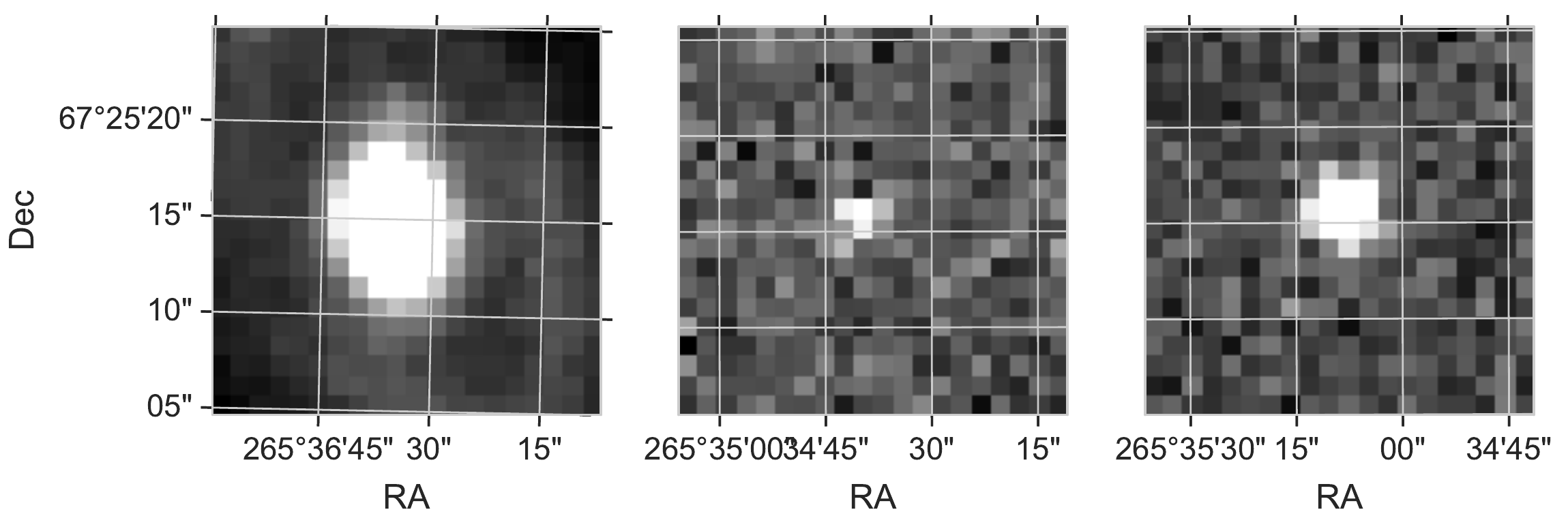}
    \caption{From left to right: $g$-band postage stamps in of the reference image (i.e. the host galaxy) centred at {\oldsn}, the difference image for \oldsn\ for the science image on July 16, 2019, and the difference image for \newsn\ centred on its position using the science image on February 7, 2020. In both cases, the SNe were close to lightcurve maximum.}  
    \label{fig:cutouts}
    \end{center}
\end{figure}

\begin{figure}
    \begin{center}
    \includegraphics[width=0.45\textwidth]{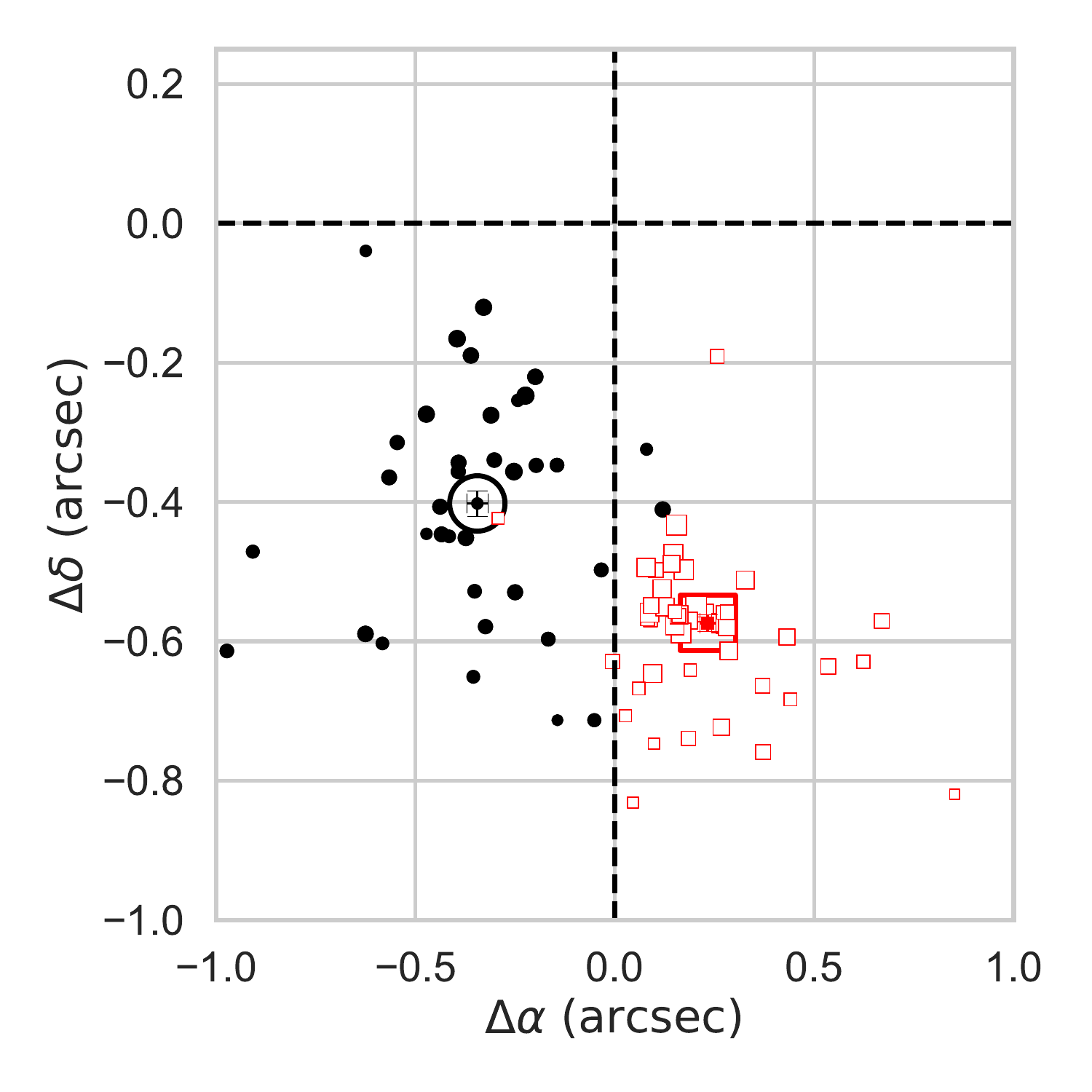}
    \caption{Positions of each alert corresponding to {\oldsn} (filled black circles) and {\newsn} (open red squares) with the size of the markers proportional to the signal to noise ratio of the alert. The median positions of the alerts corresponding to each SN are highlighted using a large black open circle for \oldsn\ , while the median position of alerts for \newsn\ is highlighted using a large open red square. The intersection of the two dashed lines shows the position of the host galaxy.}
    \label{fig:locations}
    \end{center}
\end{figure}
\begin{figure*}
    \begin{center}
	\includegraphics[width=2\columnwidth]{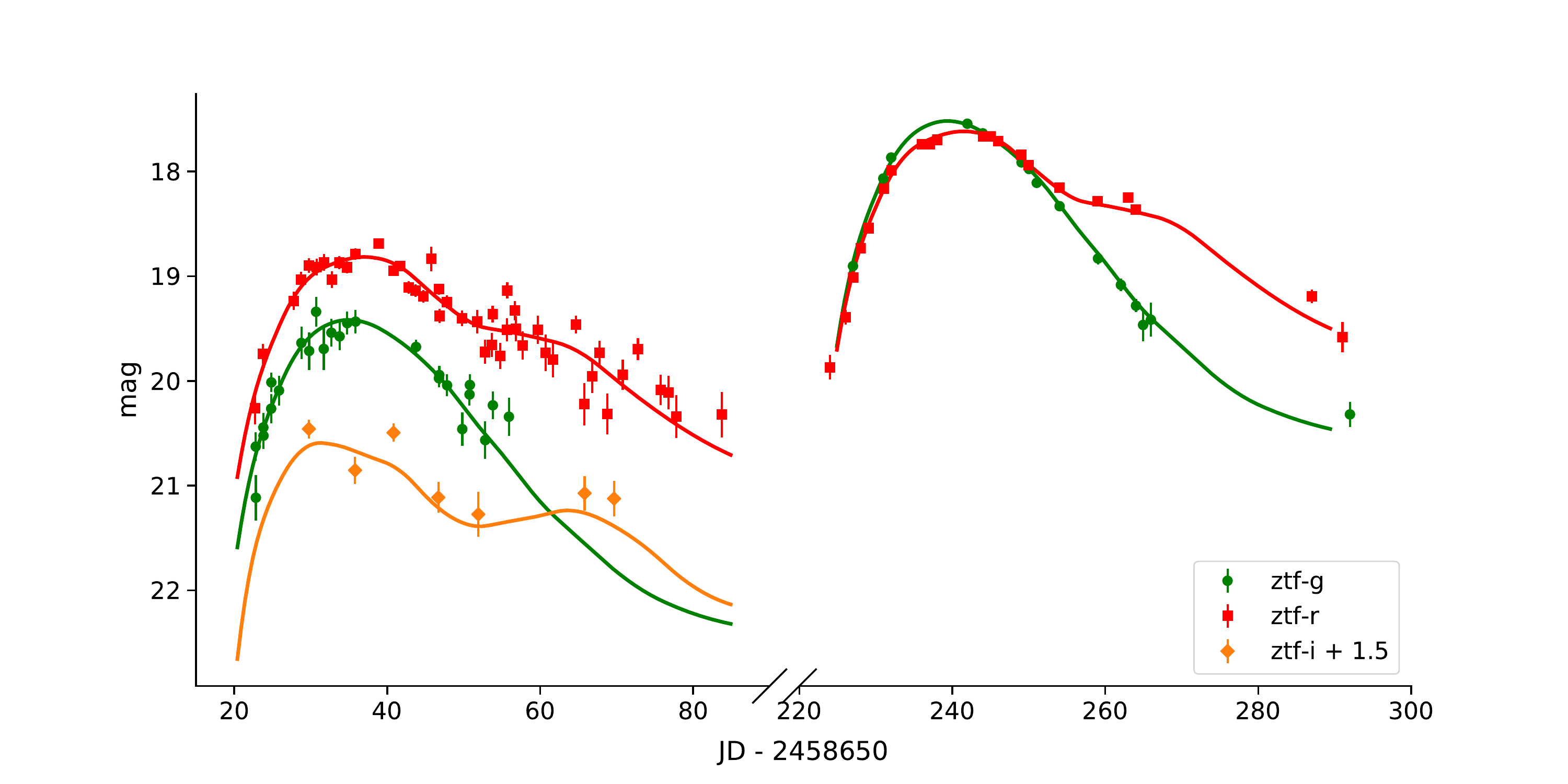}
	\caption{Observations of the sibling system at $\mathrm{SNR} > 3,$ in observed bands 
	of the two individual supernovae where forced photometry at the location of \oldsn\ after a Julian Day of 2458750 has been ignored, while early photometry at the position of \newsn\ has been ignored due to contamination. Along with the data, we show the best fit \salt\ model curves for each \snia\ . This shows the remarkable coincidence of having two \snia\ within the distance shown in Fig.~\ref{fig:locations} happen within $\sim 200\ $ days. The \salt\ parameters and uncertainties corresponding to the best fit model, and peak brightnesses in the ZTF filters are presented in Tab.~\ref{tab:app_peakmags}, and are different by $\sim 2 \ \mathrm{mag}$ in the rest frame Bessell B-band.}
	\label{fig:sibling_lc}
    \end{center}
\end{figure*}
\ztfname\ ~\citep{2019TNSTR1202....1N} is a transient detected on the core of a bright galaxy in the public 3-day cadence ZTF survey. 
It had detections in public alert photometry ($\mathrm{SNR} \ge 5$) from {\textcolor{black}{June 7 through August 14, 2019}}, and the $i$-band Partnership survey from {June 11  and Aug 23}, 2019, reaching a brightest observed magnitude of $18.69$ in the $r$-band. After that, the transient faded below detection. 
\ztfname\ was reported to TNS as \oldsn . \refcor{Since it never got as bright as $18.5$ mag, the high completeness (93 \%) threshold reported in the ZTF Bright Transient Survey~\citet{2020ApJ...904...35P}, it was, unsurprisingly, not 
 followed up spectroscopically by BTS. Unfortunately, no independent  spectroscopic classification has been reported either. }\\

On January 27, 2020, i.e., about 200 days after the detection of \oldsn, an apparent re-brightening of the source occurred, \newsn, shown in Fig.~\ref{fig:cutouts},  reaching a significantly brighter state of $17.54$ mag in $g$-band, well over the BTS classification threshold of $18.5$ mag. 
Upon closer examination, the re-brightening was not at the exact same location as the first detection. The position of the \ztfname\ alerts during this entire period are shown in Fig.~\ref{fig:locations}, with the black markers denoting epochs before Julian Day 2458800 (Nov 12, 2019), while those after this date are shown in open red squares displaying the positional clustering of alerts during these phases. Visual inspection of the light curve for the alerts during these two time periods immediately made it evident that there were two distinct explosive transients with a small projected distance between them. While we do not show this light curve (built out of alerts) in the paper, this is also apparent from the forced photometry light curve shown in Fig.~\ref{fig:sibling_lc} discussed later in this section. The separation between the two transients as shown in Fig.~\ref{fig:locations}, was $0.57 \arcsec\ .$ Consequently, these transients have been named \oldsn \ and \newsn\ .
 At the time for the second event, the field was no longer part of the 4-day cadence Partnership $i$-band survey, thus the brighter SN was only observed in $g$ and $r$-bands.\\

\begin{figure*}
    \begin{center}
	\includegraphics[trim=50 290 60 300, clip=true, width=2\columnwidth]{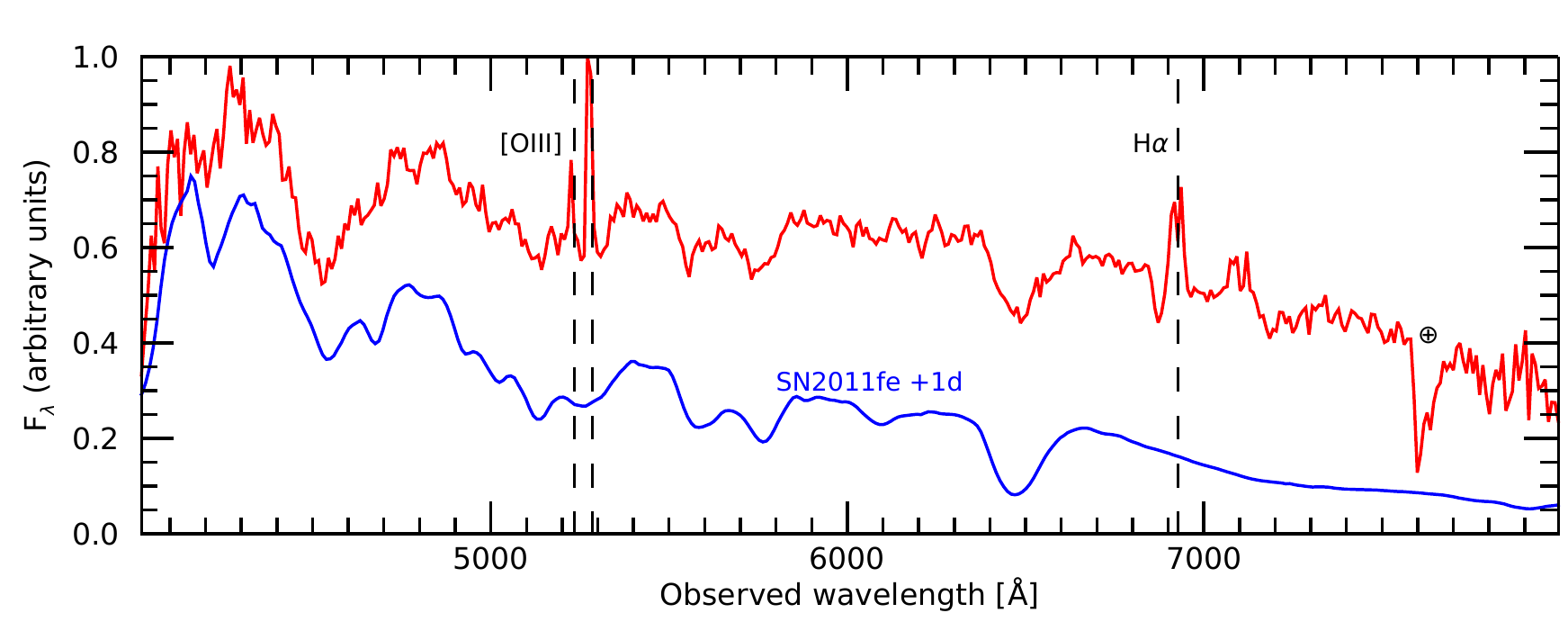}
	\caption{Spectrum of \newsn\  obtained at  
	on 2020-02-17 with the Liverpool Telescope
	. The spectrum was classified as a \snia\ in the TNS classification report~\citep{2021TNSCR.341....1P} and provides an approximate redshift of $\sim 0.055$ for the supernova. A comparison with a spectrum from SN 2011fe at the same epoch from a compilation by \citet{2014ApJ...788L..21A} is shown.
	A more accurate host galaxy redshift, $z=0.0541$, was measured based on the [OIII] and H$_{\alpha}$ lines, primarily using a spectrum from NOT, obtained more than two months later when the SN had faded.}
	\label{fig:snia_spectrum}
    \end{center}
\end{figure*}

As part of BTS, the new transient,  \newsn, was securely classified as a normal
{\snia}, \citep{2021TNSCR.341....1P} using the Spectrograph for the Rapid Acquisition of
Transients (SPRAT) \citep{2014SPIE.9147E..8HP}  on the Liverpool Telescope~\citep[henceforth LT]{2004SPIE.5489..679S}. The LT spectrum is shown in Fig.~\ref{fig:snia_spectrum}, along with a spectrum of the "typical" normal \snia\ SN 2011fe at a similar phase. However, the older transient must be photometrically classified.  We first describe the photometric light curves and then the host properties. \\

In the remainder of this section, we first discuss the association of a host galaxy with these two transients. This is followed by a description of the data processing steps needed to obtain a light curve and finally, we use both the host information and the light curves to classify the older transient.\\

\subsection{Data Processing of the Transient Light Curves}
\begin{table}
	\caption{Positions of the two supernovae {\oldsn} and {\newsn} separated by 0.57 {\arcsec} at z = 0.0541.}
\begin{tabular}{cccccccccccc}
\hline \hline
ID & RA (deg) & DEC (deg) & host sep ($\arcsec\ $) \\
\hline
{\oldsn} & 265.42935 & 67.96189 & 0.50 \\
{\newsn} & 265.42974 & 67.96183 & 0.42 \\
\hline
\end{tabular}
\label{tab:locations}
\end{table}

\begin{table*}
	\caption{Properties of the two \sne\ {\oldsn} and {\newsn}: The \salt\ parameters of the \sne\ based on a maximum likelihood fit, along with the synthetic peak magnitudes for this best fit model in the ZTF $g$-band, $r$-band and $i$-band  along with the Bessell $B$-band.}
\begin{tabular}{ccccccccc}
\hline \hline
ID & $t_0$ (day) & $10^{4} \times\ x_0$ & $x_1$ & $c$ & $g$ peak (mags) & $r$ peak (mags) & $i$ peak (mags) & $B$ peak (mags)\\
\hline
{\oldsn} & 2458685.41 $\pm\ $ 0.15 & 2.67 $\pm\ $ 0.12 & 0.54 $\pm\ $ 0.18 & 0.57 $\pm\ $ 0.04  & 19.4 & 18.8 & 19.1 & 19.7 \\
{\newsn} & 2458889.92 $\pm\ $ 0.08 & 17.08 $\pm\ $ 0.56 & 0.61 $\pm\ $ 0.13 & 0.00 $\pm\ $ 0.03 & 17.5 & 17.6 & 18.2 & 17.6 \\

\hline
\end{tabular}
\label{tab:app_peakmags}
\end{table*}

Using the alert packet from the Growth Marshal \citep{Kasliwal_2019} associated with the transient \ztfname, we split the observations into two groups assigning them to \oldsn\ if the Julian Day of the observation, $JD \leq 2458800$ (Nov 12, 2019), and \newsn\ otherwise. The time of the split was determined by visual inspection. We determined the position of each SN, by taking the median of the positions of the $5 \sigma $ detections for each SN. These locations are summarized in Table~\ref{tab:locations}, and the positions of these detections are shown in Fig.~\ref{fig:locations}. 

We run forced photometry at these SN locations using a pipeline, hereafter known as the \zuds\ pipeline\footnote{\url{https://github.com/zuds-survey/zuds-pipeline}} (Dhawan et al., in prep), which performs aperture photometry using the Astropy affiliated package  \texttt{PhotUtils}  \citep{Bradley_2019_2533376}, using a six pixel diameter aperture on the difference images. The reference images for the difference images are constructed by co-adding exposures from epochs at least 30 days or more before the initial estimate of the time of maximum from the alert photometry, using the software \texttt{SWARP} \citep{2010ascl.soft10068B}. In order to build the co-add, we only take epochs with seeing between 1.7$\arcsec$ and 3$\arcsec$ and a magnitude limit deeper than 19.2 mag. For consistency, we use the same reference image for both SNe. In the ZUDS pipeline, difference images are obtained using \texttt{HOTPANTS}~\citep{2015ascl.soft04004B}, an implementation of the image subtraction algorithm~\citep{1998ApJ...503..325A}. The zero points for each epoch are computed by IPAC, corrected for  a six-pixel diameter aperture. For the $i$-band, we use the images corrected for an observed fringing pattern, using the \texttt{fringez} software~\citep{2021arXiv210210738M}. 
From the IPAC Forced Photometry Service~\citep{2019PASP..131a8003M} at the same locations, we obtain the metadata for each observation, including  the magnitude limit $m_\mathrm{lim}$ of the observation, the seeing $\mathrm{seeing}$ of the observation and, the standard deviation $\sigma_{\mathrm{pix}}$ on the background at the pixel on which the SN is located. We combine this information with the \zuds\ pipeline results for data quality assessment. 
Specifically, we only use those observations that satisfy the following conditions:
$
1.0\arcsec < \mathrm{seeing} < 4.0\arcsec, \quad
m_{\mathrm{lim}} < 19.2 \mathrm{\, mag}, \quad 
\sigma_{\mathrm{pix}} < 14.0,
$
where $\sigma_{\mathrm{pix}}$ is the robust sigma per pixel in the science image and is used as a metric to remove non-photometric data. We then use a maximum-likelihood method to fit the \salt\ model to each of these two supernova light-curves. The low seeing values are removed to protect against undersampling during image subtraction. We then remove the epochs that have $5-\sigma$  flux outliers relative to the best fit \salt\ model (discussed later and summarized in Tab.~\ref{tab:app_peakmags}) and use the remaining selected points as the light curves of the individual supernovae. The final photometry datasets used are included as Tables \ref{tab:oldsn} and \ref{tab:newsn}.
The resulting light curves are shown in Fig.~\ref{fig:sibling_lc}. As stated earlier, this light curve clearly shows the presence of two transients with no detections for a period of over 100 days in between. For the unclassified transient \oldsn\ , 
we notice that \oldsn\ has a shoulder in the redder bands ($r-$band and $i-$band) as seen in Fig.~\ref{fig:sibling_lc} strongly indicating that the \sn\ is of Type Ia. We will verify this shortly after discussing its redshifts and host properties. \\

\begin{figure}
    \begin{center}
    \includegraphics[width=0.45\textwidth]{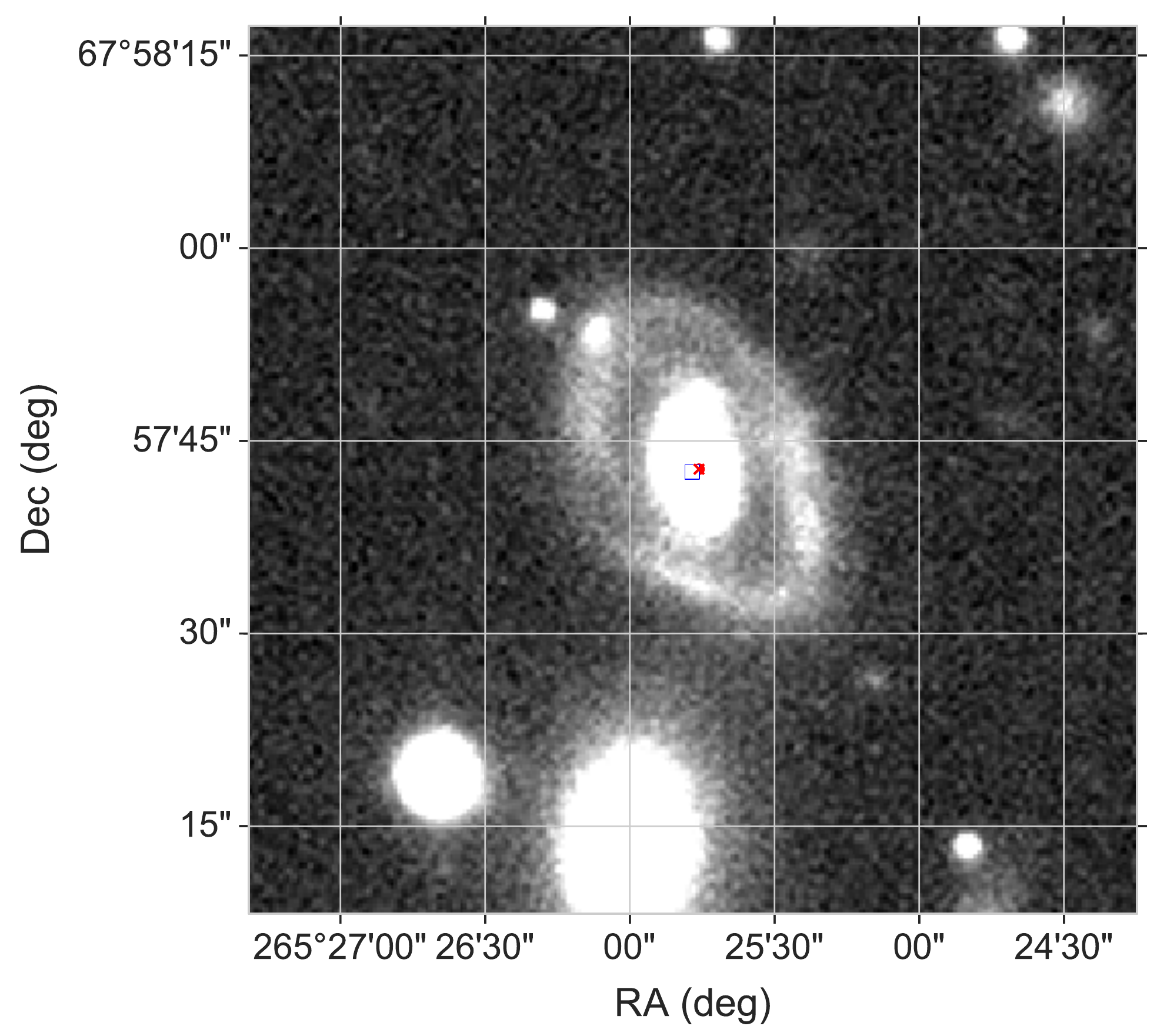}
    \includegraphics[width=0.45\textwidth]{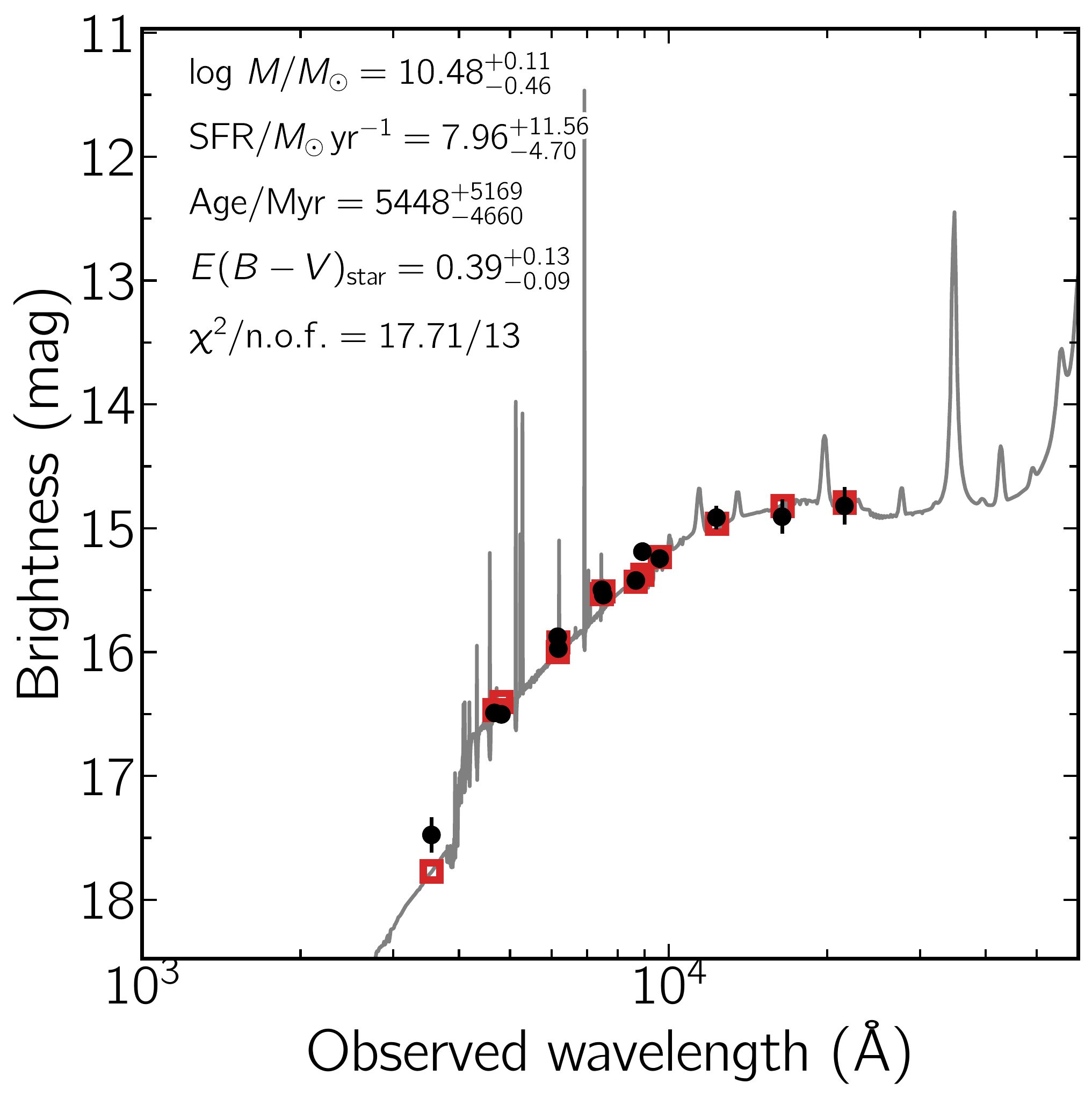}
\caption{Host association and Properties: (Top panel) Location of nearby galaxies obtained from the Legacy Survey
	in g band from the NOAO data lab \changes{are shown along with the supernovae. The location of \oldsn\ is indicated with} a red cross, while \newsn\ is shown with a blue open square almost
spatially coincident in the core of the host galaxy
(Bottom Panel) The best fit model spectral energy distribution of the host galaxy of ZTF19aambfxc (photometric observations {$\bullet$}, 
model-predicted magnitudes {\textcolor{red}{$\square$}}) from multi-band photometry. The solid line displays the best-fitting model of the SED. The fitting parameters are shown in the upper-left corner. The abbreviation n.o.f. stands for numbers of filters.}
    \label{fig:host}
    \end{center}
\end{figure}

\textbf{Host Association and Properties:} 
The top panel of Fig.~\ref{fig:host} shows the positions of the supernovae and the two nearest galaxies. Both supernovae are found to lie on the core of a galaxy, while there is a second nearby galaxy, approximately $30 \arcsec\ $ away which we deem extremely unlikely to be the host of any of the two SNe. We will refer to the first galaxy as the host galaxy. A high SNR spectrum from the ALFOSC instrument on the Nordic Optical Telescope (NOT) \footnote{PI: Sollerman \& Goobar} taken on April 28, 2020 was used to determine the properties of the galaxy. The host galaxy redshift was accurately measured to be $z=0.0541$ from the position of [OIII] and H${\alpha}$ lines, in excellent agreement with the best fit to the SN spectrum of \newsn\ $\sim 0.055.$ This late spectrum is available from the authors upon request. \refcor{The angular distances along with the absence of a nearby galaxy establishes the two transients as siblings, i.e. have the same host galaxy. Additionally, the spectrum from the host provides us with an accurate measurement of their common redshift, used in the calculations later.}\\

While the two conclusions above are central to this paper, we expand our study of the properties of the host galaxy as it might be relevant to the conclusions on extinction.  We retrieve the science-ready co-added images from the Sloan Digital Sky Survey data release 9 (SDSS DR 9; \citealt{Ahn2012a}), the Panoramic Survey Telescope and Rapid Response System (Pan-STARRS, PS1) DR1 \citep{Chambers2016a}, the Two Micron All Sky Survey \citep[2MASS;][]{Skrutskie2006a}, and preprocessed \wise\ images \citep{Wright2010a} from the unWISE archive \citep{Lang2014a}\footnote{\href{http://unwise.me}{http://unwise.me}}. The unWISE images are based on the public \wise\ data and include images from the ongoing NEOWISE-Reactivation mission R3 \citep{Mainzer2014a, Meisner2017a}.
We use the software package LAMBDAR (Lambda Adaptive Multi-Band Deblending Algorithm in R) \citep{Wright2016a}, which is based on a software package written by \citet{Bourne2012a} and tools presented in \citet{Schulze2020a}, to measure the brightness of the host galaxy. The spectral energy distribution (SED) was modelled with the software package Prospector\footnote{\href{https://github.com/bd-j/prospector}{https://github.com/bd-j/prospector}} version 0.3 \citep{Leja2017a}. We assumed a linear-exponential star-formation history, the \citet{Chabrier2003a} IMF, the \citet{Calzetti2000a} attenuation model, and the \citet{Byler2017a} model for the ionized gas contribution. The priors were set as described in \citet{Schulze2020a}.\\

The best fit host galaxy SED, along with its parameters, is shown in the lower panel of Fig.~\ref{fig:host}. 
This shows that the stellar mass of the galaxy is $\sim 10^{10.48^{+0.11}_{-0.46}} {M_\odot}.$ This is approximately equal to the threshold usually chosen to divide \snia\ samples into two groups to which different corrections are applied. Fortunately, this will not have an impact on our calculations which depend on the difference of the distance moduli of the pair as the pair share the same host galaxy. We also note that the measured stellar mass is very comparable to the Milky-Way value reported by \citet{2015ApJ...806...96L}, thus the dust properties of this host galaxy are expected to be similar to the Milky Way, even in a model like ~\citet{2020arXiv200410206B}.\\

\begin{table}
    \centering
    \caption{Bayesian Information Criterion of different SN types fitted to the multi-band photometry of \oldsn\, shown in Fig.~\ref{fig:sibling_lc}.}
    \begin{tabular}{|c|c|c| }
    \hline
    SN type & BIC & $\Delta \mathrm{BIC} $\\
    \hline
    SN~Ia: Norm   &  419.47 & 0.\\
    SN~Ia: 91bg     & 1200.22 & 780.75 \\
    SN~Ibc & 1042.55 & 623.08\\
    \hline
    \end{tabular}
    
    \label{tab:bic_sntypes}
\end{table}

\textbf{Photometric Classification of the \oldsn\ :}
To verify the \snia\ class suspected from the light curve shape, we \changes{classify the type from the photometry. Aside from the light curve shape, we note that the supernova is 2 mags dimmer than \revtwo{expected of a SN Ia} in the same galaxy \revtwo{as demonstrated by \newsn }.  Additionally the stretch $x_1$ of the {\oldsn} is found to be $0.54 \pm 0.18.$  Thus we limit the comparison to models of dimmer supernovae like core-collapse supernovae and peculiar supernovae like 1991bg-like SNe, while  excluding over-luminous peculiar types like 1991T on the basis of the peak brightnesses and 
the stretch.}
The maximum likelihood fit to to the lightcurve in Fig. \ref{fig:sibling_lc} with models of different supernova classes and sub-types. 
Since the models considered have different numbers of parameters and are non-nested, we cannot use the likelihood ratio test to compare the models. Rather than computing the Bayesian evidence for each model which is computationally expensive, and  the population priors are likely not well known, we use the Bayesian Information Criteria (BIC)~\citep{1978AnSta...6..461S} for each model, which for Gaussian distributed deviations amounts to
\be
\mathrm{BIC} = - 2 \log{(L_\mathrm{max})} + k \log{(n)} 
\ee
where $L_\mathrm{max}$ is the maximum of the log likelihood, $n$ is the number of data points and $k$ is the number of parameters in the model. The difference in BIC values between two models is approximately proportional to the logarithm of the Bayes ratio of the two models, with $\Delta BIC = 10 $ considered decisive evidence against the model with higher BIC~\citep{2007MNRAS.377L..74L}.  
 Applying a maximum likelihood fit to different kinds of templates of supernovae types, we calculate the BIC values as shown in Table~\ref{tab:bic_sntypes}. 
 We choose the SN Ib/c template kindly provided Peter Nugent~\footnote{{\url{https://c3.lbl.gov/nugent/nugent\_templates.html}}} and the K-correction estimates in \citet{2002PASP..114..803N}. Since the observed colours of the SN are red, we also fit the SN1991bg template from \citet{2002PASP..114..803N}.  The column $\Delta BIC$ has values of $\mathrm{BIC}(model) - BIC(\salt).$ Thus models considered here with $\Delta BIC >> 10$ are decisively disfavoured. Thus, the entries in Table~\ref{tab:bic_sntypes} show that from the photometric data alone, we can say that of the types of supernovae and templates considered, {\oldsn} is a Normal \snia\ .\\

\section{Method}
\label{sec:methods}
Sibling supernovae are inferred to be in the same galaxy through their transverse proximity, and thus have virtually identical (radial) distance. First, we quantify the potential difference in the radial distance in comparison to the intrinsic dispersion of \snia. Galaxies are typically of size $\sim \kpc\ ,$ and host \snia\ within a few tens of \kpc\ from the centre of the galaxy~\citep{2012ApJ...755..125G,2020arXiv200809630G}. 
Thus, sibling supernovae are expected to have a distance difference $\delta d \lesssim 100 \ \kpc\ ,$ resulting in a distance modulus difference of $5/\ln{(10)} \times \frac{\delta d}{d}.$ For a sibling at a redshift of $z = 0.0541,$ even for a distance difference of $\sim 100 \ \kpc\,$ this difference is $\approx 10^{-3}$ which is two orders of magnitude smaller than the distance uncertainty induced by the intrinsic dispersion of $\sigma_{int} \sim 0.1 \ $  of \snia\ . This understanding can be expressed as a prior probability which is a normal distribution in the difference of the distance moduli with a standard deviation related to the intrinsic dispersion
\be
\Pi(\Delta \mu \vert \mathcal{H}) =  \mathcal{N}(\Delta \mu, \mathrm{\sigma}^2).
\label{eqn:prior_dist}
\ee

In this work, we assume the \salt\ model with an intrinsic dispersion affecting the brightness of \snia\ coherently at all wavelengths (technically following the G10 intrinsic dispersion~\citep{2010A&A...523A...7G}). This means that the estimated distance modulus of a supernova can be normally distributed about the true distance modulus with a standard deviation of $\sigma_{int}.$ The value of the intrinsic dispersion in supernova samples is determined for each sample, and is generally $\sim 0.1 \ \mathrm{mag}\ .$ The intrinsic dispersion of \snia\ in the same galaxy (or equivalently if sibling supernovae have correlated intrinsic scatter) has been investigated
~\citep{2020ApJ...896L..13S,2020ApJ...895..118B} and found to be consistent with low correlations $r$. Thus, we choose the standard deviation in our prior probability to be $\mathrm{\sigma}^2 = 2\sigma_{int}^2 (1- r),$ with a fiducial value of $r=0.\ $ We also use the \changes{SALT standardization relation}~\citep{1999ApJ...525..209T} as used in recent supernova cosmology analyses ~\citep{2019ApJ...874..150B,2019ApJ...876...15H,2018ApJ...859..101S,2018ApJ...857...51J,2014A&A...568A..22B}
\be
\mu_i = {m_B^\star\ }_i + \alpha \cdot {x_1}_i- \beta \cdot c_i - {M_B^\star} + \delta_{\mathrm{Host_{i}, loc_{i}}} \mu,
\ee
where, for completeness, we have included a correction term for (potentially local) environment dependence.  The term involving the impact of the host galaxy has been mostly used in the supernova cosmology literature as a step function involving the global properties of the host galaxy such as stellar mass $M_\mathrm{stellar}^\mathrm{Host},$  as in
$
\delta_{\mathrm{Host, loc}} \mu = \mathrm{step} \times (M_\mathrm{stellar}^\mathrm{Host} - M_\mathrm{thresh}) 
$
where $\mathrm{step} \sim 0.1\ \mathrm{mag}\ $ and $M_\mathrm{thresh} \sim 1 \times 10^{10} M_\odot $ are obtained by minimizing the Hubble residuals from a fiducial model. For such a model where the environmental dependence is through global properties of the host, clearly this difference disappears for supernova siblings. However, such environmental dependence is expected to depend on the local properties of the galaxy. For example, this may be driven by the properties of the progenitor(s) which inherited the properties of the local stellar population or due to dust which could also be local. This could have important consequences for the measurements of H$_0$ \citep{2015ApJ...802...20R,2020A&A...644A.176R}, though the details are part of a current debate~\citep{2018ApJ...867..108J}. Thus, these siblings, even though separated by a tiny projected distance, could be further apart in the galaxy (radially) and have different properties. Nevertheless, as far as estimated corrections go, corrections from local properties are made by analyses restricted by projected distances, often in regions of projected distances of  $\sim 2 \ \kpc .\ $The siblings being \changes{$0.6 \ \kpc\ $ (using a Planck cosmology~\citep{2020A&A...641A...6P})} apart are close enough that any correction term due to local measurements would also be extremely similar for the two \sneia\ and thus the difference would be small. Thus, we set $\delta_{\mathrm{Host, loc}}$ to $0$ in the rest of this work. \\

Thus, for a pair of sibling supernovae, we get
\be
\Delta \mu \equiv
\mu_1 - \mu_2 = -2.5\log{({x_0}_1/{x_0}_2)} + \alpha \cdot ({x_1}_1 - {x_1}_2)- \beta \cdot (c_1 - c_2)
	      \label{eqn:dist_diff}
\ee
where we have used the approximation $m_B^\star\ = - 2.5 \times \log10(x_0) + K,$ where K is a constant. 

In this model, the parameters are $\Psi = \left\{\psi, \phi_{1}, \phi_{2} \right\},$ where the \salt\ model parameters for each supernova are
$\phi \equiv \{ x_0 ,\ x_1 ,\ c \},$ and the subscripts $1$ and $2$ refer to the two supernovae, and $\psi \equiv \{\alpha, \beta\}.$ We can 
write the posterior distribution \refcor{$P(\Psi \vert D, \mathcal{H})$ on $\Psi$} where $D$ is the photometric data, along with spectroscopic data needed to determine the redshift $z,$ and $\mathcal{H}$  
represents our understanding of the astrophysics leading to the equations we use: 
\begin{eqnarray}
	P(\Psi \vert D, \mathcal{H}) &\propto& P(D \vert \Psi, \mathcal{H}) \Pi(\Psi \vert \mathcal{H}) \nonumber\\
				     &=& P(d_1 \vert \phi_1, \mathcal{H}) P(d_2 \vert \phi_2, \mathcal{H}) \Pi(\Psi \vert \mathcal{H}) \nonumber\\
	\Pi (\Psi \vert \mathcal{H}) &=&  (\mathcal{N}(\mu_1 - \mu_2, \mathrm{\sigma}^2)  \Pi(\alpha) \Pi(\{\beta, \phi_1\, \phi_2\ \}),
\label{eqn:posterior}
\end{eqnarray}
where $P(D \vert \Psi , \mathcal{H})$ is the likelihood function of $\Psi,$ and $\Pi(\Psi \vert \mathcal{H})$ is the prior on $\Psi.$ Utilizing the fact that the time of peak of the supernovae are reasonably well constrained, and the flux due of \oldsn\ is negligible at the time of \newsn\ which peaks $205$ days after \oldsn\ , factorizing the likelihood function into independent likelihood functions for the data of \oldsn\ and \newsn\ is an excellent approximation. Therefore, we write this as the product of individual likelihood functions 
$P(d_i \vert \phi_i, \mathcal{H})$ is the likelihood function of the \salt\ model parameters for the $\mathrm{i th}$ supernova. \changes{This model for the supernova photometry includes the impact of Milky Way extinction by multiplying the SALT spectral energy distributions by a time independent, wavelength dependent extinction calculated using the fitting functions of ~\citet{1989ApJ...345..245C} with the EBV value for Milky Way at the location of the supernovae evaluated using the recalibration~\citep{2011ApJ...737..103S} of the~\citet{1998ApJ...500..525S} dustmaps.}  Finally the likelihood function encodes the assumption that the measured fluxes are Gaussian distributed about the \salt\ model fluxes with variances described by the flux uncertainties reported in the photometry, as well as the uncertainties on the model as determined from \salt\ training. The prior $\Pi (\Psi \vert \mathcal{H})$ includes chosen priors on each of the parameters $\Pi(\alpha) \Pi(\beta\,\phi_1\,\phi_2)$. Our chosen priors on each parameter in $\Psi$ in this paper are un-informative (uniform, hard priors) except for $\alpha$ for which we sometimes adopt the Pantheon result~\citep{2018ApJ...859..101S}, as described as part of each calculation in Sec.~\ref{sec:results}.
Aside from these choices, we use the result of a different measurement (galaxy association) affirming that these \snia\ are siblings, to further constrain these parameters, expressed as the normal distribution~Eqn.~\ref{eqn:prior_dist} using Eqn.~\ref{eqn:dist_diff} for $\Delta \mu$. We use \texttt{emcee}~\citep{2013PASP..125..306F} to explore the parameter space.
\\

\section{results}
\label{sec:results}

\begin{figure*}
    \begin{center}
    \includegraphics[width=2.15\columnwidth]{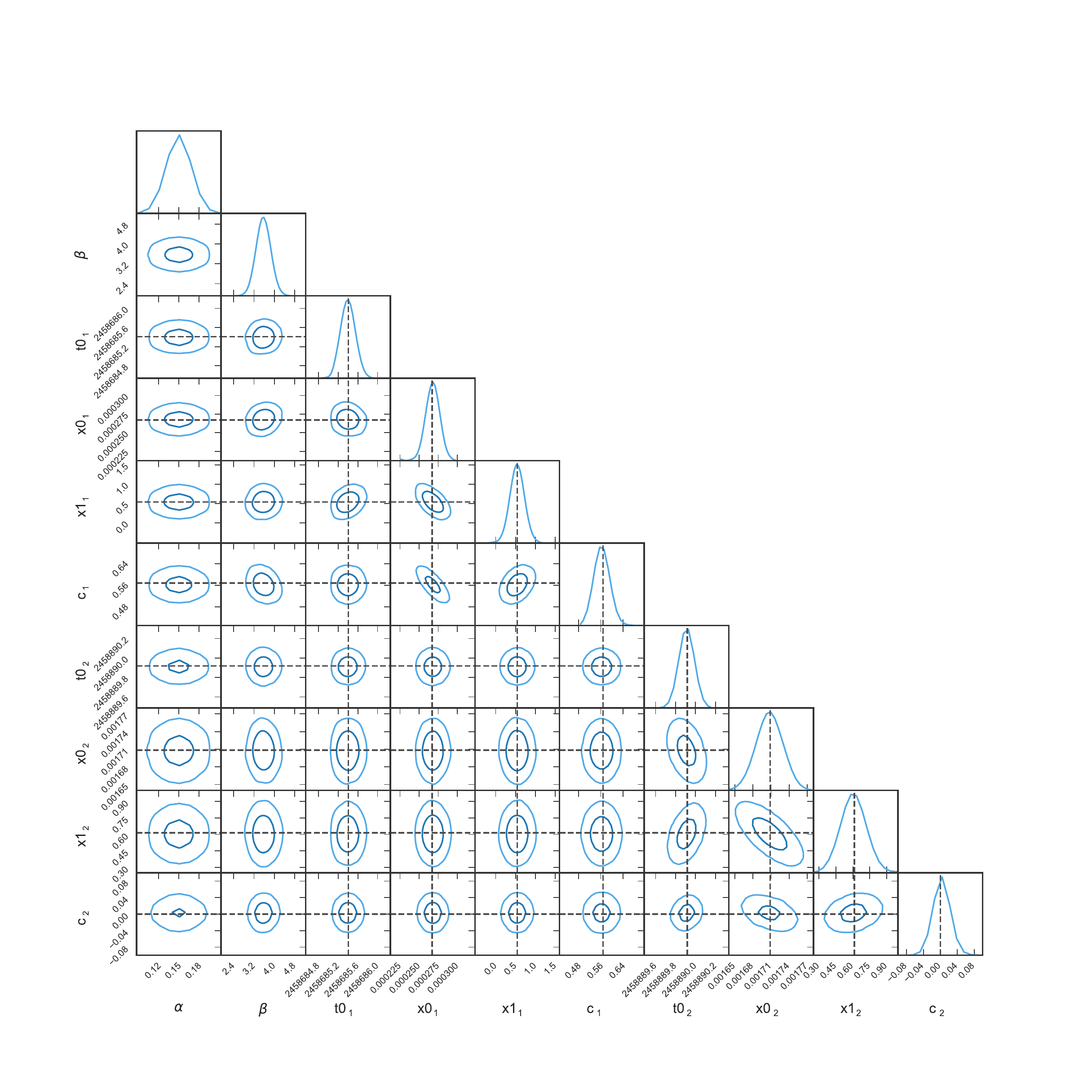}
    \caption{The joint posterior distribution of the \salt\ light curve parameters for the two supernovae and the global parameters $\alpha$ and $\beta.$  The contours enclose 68\% and 95\% of the probability while the dashed lines show the maximum likelihood estimates of the parameters shown in Tab.~\ref{tab:app_peakmags} if available from each of the single supernovae. Uninformative uniform box priors were used as hard priors on all of the parameters except $\alpha$ where a Gaussian prior of $0.15 \pm 0.01\ $ was used incorporating the values obtained in the analysis of the Pantheon data set~\citep{2018ApJ...859..101S}.}
    \label{fig:joint_posteriors}
    \end{center}
\end{figure*}

\begin{figure}
    \begin{center}
    \includegraphics[width=0.45\textwidth]{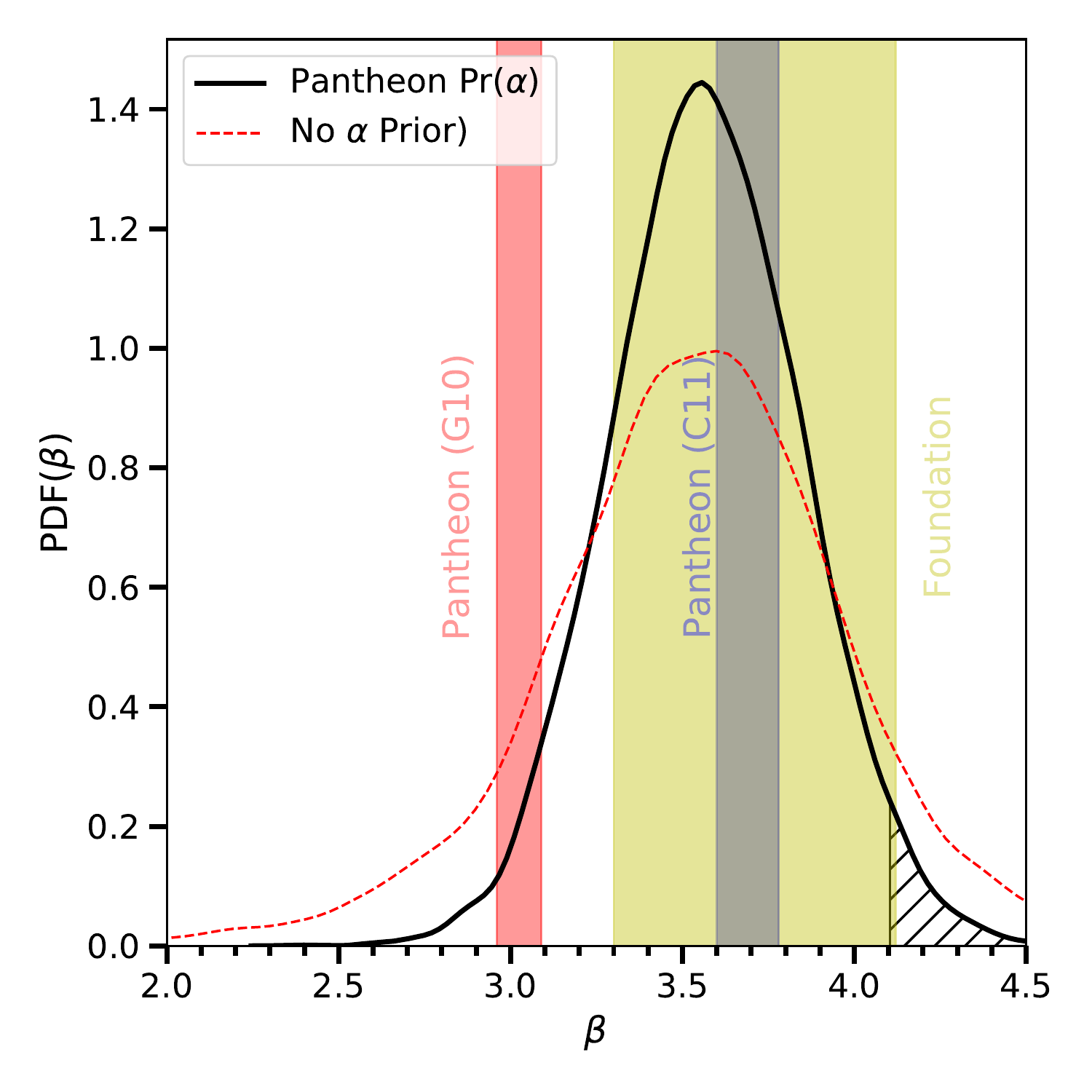}
    \caption{The PDF of $\beta$ marginalized over all other model parameters is shown in solid (dashed) black (red) curve for an intrinsic scatter of 0.1 and correlation r=0, with (without) the Pantheon priors on $\alpha.$  The constraints are $\beta =$ 3.4(3.6)  $\pm$ 0.3 (0.4). The shaded vertical rectangles show the $1-\sigma$ constraint on $\beta$ from the PanSTARRS spectroscopic sample~\citep{2018ApJ...859..101S} for the G10 model of intrinsic scatter (red) and the C11 model (blue) and for 69 low redshift \snia\ from the Foundation Supernova Survey (yellow)~\citep{2021arXiv210206524D}. The hatched \changes{black} region shows the probability of the $\beta$ value being at least as large as the value expected if the colour dependent extinction was solely due to ISM host dust of the same nature as the Milky Way. This probability associated with this region is 2.5 {\%}.}
    \label{fig:pdf_beta}
    \end{center}
\end{figure}

\begin{figure}
    \begin{center}
    \includegraphics[width=0.45\textwidth]{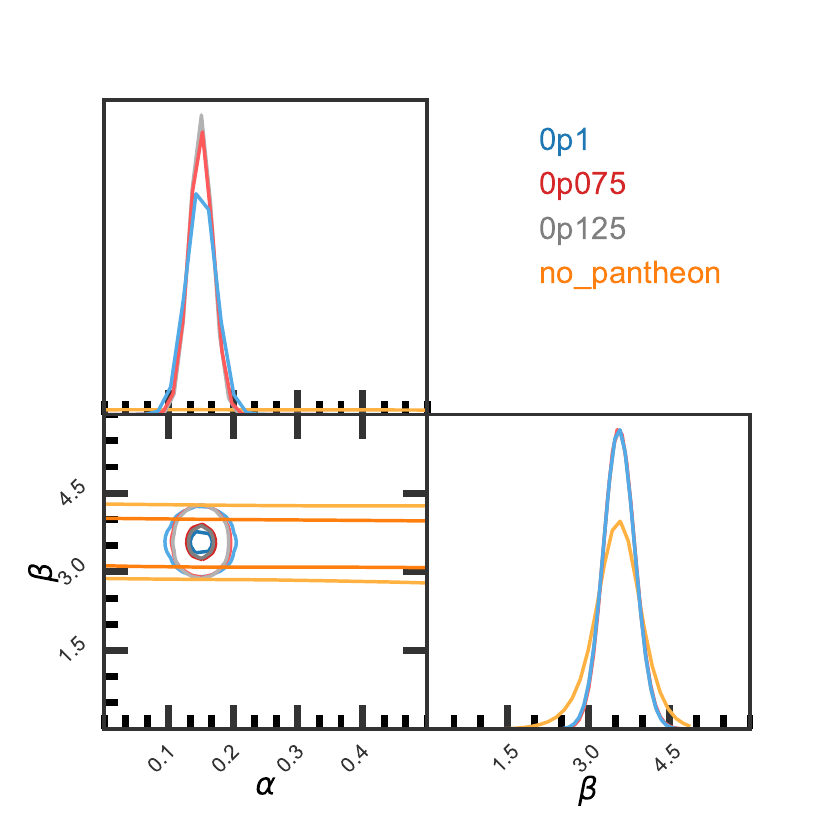}
    \caption{Joint Constraints on $\alpha$ and $\beta$ from the \snia\ sibling system when the magnitude of intrinsic dispersion used in calculating the posterior is varied from the fiducial value of $0.1$ (blue) to $0.075$ (red) and $0.125$ \changes{(gray)}. Finally, the orange contours show the results for the fiducial choice of $\sigma_{int} = 0.1$ when the Pantheon priors on $\alpha$ are not used. This leaves $\alpha$ completely unconstrained. but the constraints on $\beta$ only change slightly from $3.4 \pm 0.3$ to $3.6 \pm 0.44$}
    \label{fig:prior_impact}
    \end{center}
\end{figure}
\textbf{Constraints on $\beta$: } Having confirmed that the siblings are \sneia\, we employ the methodology discussed in Sec.~\ref{sec:methods} to calculate the posterior distribution of all the parameters $\Psi$ based on the data from the siblings in Fig.~\ref{fig:joint_posteriors}. We take into account the spatial coincidence of this sibling pair, thereby using Eqn.~\ref{eqn:dist_diff} and calculate the posteriors using Eqn.~\ref{eqn:posterior}. In Fig.~\ref{fig:joint_posteriors}, the blue contours enclose 68\% and 95\% of the probability while the dashed black lines show the maximum likelihood estimates of the parameters from each of the single supernovae. Uninformative uniform box priors were used as hard priors on all of the parameters except $\alpha$ where a Gaussian prior of $0.15 \pm 0.01\ $ incorporating the values obtained in the analysis of the Pantheon data set~\citep{2018ApJ...859..101S} were used. The constraints on the parameter $\alpha$ effectively stems entirely from the Pantheon prior. The excellent match between the maximum likelihood estimates of the individual supernovae in Tab.~\ref{tab:app_peakmags}~(with no knowledge of the sibling nature) and the posteriors of the joint likelihood confirm the expected result that \salt\ parameters of individual supernovae are not affected by the global parameters like $\alpha$ and $\beta$. Thus, the only new information from the sibling nature is the constraints on $\beta$ which would be entirely unconstrained from the individual fits of two supernovae. \\

\textbf{Consistency of $\beta$ constraints with previously reported values:} Therefore, we focus on the constraints on $\beta$ from this system. In Fig.~\ref{fig:pdf_beta}, we show the posterior probability density function on $\beta$ inferred from the sibling supernova system, when marginalized over all other parameters in $\Psi,$ and using the Pantheon prior on $\alpha.$ The constraints are $3.5 \pm 0.3$ which is a $8 \%\ $ measurement on the parameter $\beta$ from this system alone. This can be compared to constraints on $\beta$ from cosmological surveys, in terms of consistency of values summarized in Tab.~\ref{tab:app_peakmags} and the magnitude of uncertainty. Supernova cosmology survey results in the past decade using the \salt\ model include the Joint Light Curve Analysis (JLA)~\citep{2014A&A...568A..22B}, the results on the Pantheon sample~\citep{2018ApJ...859..101S,2018ApJ...857...51J} from PanSTARRS and the Dark Energy Survey~\citep{2019ApJ...874..150B}. All of these surveys presented results with the G10 and the C11 intrinsic scatter models. Of these the JLA studied the constraints on $\alpha, \beta $ based on different cosmological models, and the use of complementary information from Cosmic Microwave Background and galaxy surveys, while the other studies presented cosmological model insensitive constraints using \texttt{SALT2mu}. For the G10 scatter model, JLA reported values of $\beta$ between $3.099 - 3.126 \ $ with an uncertainty of $\sim 0.075 - 0.1$ while the DES/PanSTARRS analysis obtains a slightly lower value $3.02-3.03$ with an uncertainty of $\sim 0.11-0.13$. The values with a C11 intrinsic scatter model based on the SNFactory studies 
are consistently higher, ranging from $3.27-3.4$ with a similar uncertainty $\sim 0.1$ while the values obtained from DES and PanSTARRS range from $3.51-3.61$ with the uncertainties of the order $0.15-0.25.$ Clearly the values here are very consistent with C11 intrinsic scatter model values reported from past surveys, while the the consistency with G10 scatter model based analyses is at the level of $ < 2 \sigma.$ Finally, we show the comparison with a determination of $\beta$ from 69 low redshift \snia\ ~\citep{2021arXiv210206524D} from the Foundation Supernova Survey~\citep{2018MNRAS.475..193F}.\\

\textbf{Impact of Priors and Assumptions:}
In this work, we do not estimate the intrinsic dispersion as is customary in supernova cosmology analysis. Instead, we posit that the magnitude of intrinsic dispersion is $0.1$ based on results from previous surveys. Therefore, in Fig.~\ref{fig:prior_impact}, we study the impact of constraints on $\alpha$ and $\beta$ as we vary the magnitude of the intrinsic dispersion. Fig.~\ref{fig:prior_impact} shows the constraints on $\alpha$ and $\beta$ for the fiducial case (shown in Fig.~\ref{fig:joint_posteriors}) of $\sigma_{int} = 0.1 $ (blue) and two additional cases: $\sigma_{int} = 0.075$ (red) and $\sigma_{int} = 0.125$ \changes{(gray)}. Within this small range of changes to the assumed values of intrinsic dispersion, the constraints on $\beta$ are unaffected. Finally, this figure also explores the impact of not using the Pantheon data set to put a prior on $\alpha$ (orange). We find that removing this prior leaves 
$\alpha$ entirely unconstrained but the constraints on $\beta$ only change modestly from $\beta = 3.4 \pm 0.3$ to $3.6 \pm 0.4.\ $ We can understand this insensitivity as arising from the peculiarities of the sibling system: The $x_1$ values of the two \snia\ have virtually the same value, while the difference in $c$ is large. This leads to a very tiny dependence on $\alpha$ in Eqn.~\ref{eqn:dist_diff}, and consequently has small effects on the constraints on $\beta.$ The downside of this same peculiarity is that  without priors, the constraints on $\alpha$ are extremely weak as shown in Fig.~\ref{fig:prior_impact}.\\

\textbf{Comparison with Milky Way $R_B$: } Data driven models like \salt\ do not require any physical interpretation to parameters like $\beta,$ but as a best fitted value to the \changes{SALT standardization relation} for a sample of \sneia.\ However, the colour excess (after correction for Milky Way reddening) is interpreted in other models (see for example,~\citet{1999ApJ...525..209T}) as dust-extinction in the host galaxy and is described by a fitting function dependent on the total-to-selective extinction ratio $R_V$ \citep{1989ApJ...345..245C}, or in the restframe $B$-band, $R_B=R_V+1$. 
In such models, one cannot exclude different values of $R_V$ in each galaxy, describing the size and composition of the dust-grain population in the host galaxy ISM along the line-of-sight. 
Such a measurement of $R_V$ in diverse populations of galaxies is poorly known, but is well measured in the Milky Way, where it varies with direction and has an average value of $R_V=3.1$. Using the \citet{1998A&A...331..815T} relation, one may expect that the value $\beta$ is related to the $R_V$ for extinction as $\beta \approx R_B$. Supernova surveys (as discussed above) usually obtain values of $\beta$ consistent with $R_V\approx 2$, well below the average Galactic value, but cannot probe the distribution or variance of $R_V$. This can be done using longer lever arms in colour \citep[see e.g.][and references therein]{2015MNRAS.453.3300A}. In this work we show that one can also extract it from galaxies that host supernova siblings, as the $\beta$ value inferred here is specific to the host galaxy.  The $R_V$ value implied by our constraints is similar $\sim 2.5,$ and the probability of the implied $R_V$ value being at least as large as the average Milky Way value has a total associated probability of  $2.5 \%$. We note, however, that deviations from the "standard" dust law have also been reported in Galactic studies \citep[see e.g.,][]{2016MNRAS.456.2692N}. \\

Since this determination of $R_V$ is model dependent, we study this in other models as well. First, as a consistency check, we also compute the total-to-selective absorption ratio, $R_V$, using the SN~Ia model in \texttt{SNooPy} \citep{2011AJ....141...19B}. \texttt{SNooPy} fits the host galaxy $R_V$ and colour excess, $E(B-V)$ using templates based on the colour-stretch parameter, $S_{BV}$ \citep{2014ApJ...789...32B}. We present details of the inference procedure in Appendix~\ref{sec:app_snoopy}, but summarize the results $R_V \sim 2.8.\ $ Finally, we also look at the more general dust model from ~\citet{2020arXiv200410206B} in Appendix~\ref{app:bs}. In this model, the colour of a \snia\ as determined by \salt\ is postulated to be a linear combination of reddening due to  to dust in the host galaxy, and colour of the supernova with respect to a template with coefficients $R_B$ and $\beta.$ Thus, there are two colour laws in this model, and more free parameters compared to the traditional \salt\ model. The parameter $\beta$ in this model is not directly related to $R_B.$ While we postpone a more complete uncertainty analysis to future work, we show that the sibling system can be used to probe such models. For example, it suggests that if $R_V$ is 3.1 (the host galaxy of these siblings has a similar stellar mass as the Milky Way), we can show that this suggest a difference in $c_{int}$ of $\gtrsim 3$, which is much larger than the standard deviation of the Gaussian distribution for individual SN, which is ameliorated for a lower $R_V$ and low values of $\beta$. Therefore, in multiple models, this sibling pair suggests that the host galaxy has a lower value of $R_V$ than the Milky Way.\\

\textbf{Uncertainty and Future Prospects:}
As discussed earlier, the uncertainty on $\beta$ obtained from cosmological surveys in the last decade depend on the intrinsic scatter model used, and range from $0.075 - 0.15$ for the G10 model, and $0.15-0.25$ using the C11 model. These surveys use a large number of \snia\, for example $740\ $ in JLA, $207\ $ in DES, $1048$ in the Pantheon Sample, $1364$ in the photometric PanSTARRS analysis to obtain such constraints and are tighter \changes{than} the constraints from this single system by a factor of $2-3.$  For constraints from sibling supernovae to be useful in understanding distance modulus bias, it is important  for these constraints to be competitive, which is possible by combining the results from a number of \snia\ siblings. Currently, ZTF has a few \snia\ siblings (see Graham et al., 2021, in prep), while data collection continues in phase II. DES has published a few siblings, and one can expect more siblings from large ongoing surveys like ZTF, Young Supernova Experiment~\citep{2020arXiv201009724J} and upcoming surveys like 
the Legacy Survey of Space and Time~\citep{2009arXiv0912.0201L,2019ApJ...873..111I} at the Vera C. Rubin Observatory. While a simple square root of N calculation would suggest that $\sim 10 $ sibling systems should tighten the constraints to competitive levels, this actually depends on the properties of the \snia\ themselves. \\

\section{Discussion}

\changes{In this paper, we have exploited the fact that the distance moduli of two supernovae in the same galaxy are virtually identical 
and any radial difference is much smaller than the impact from intrinsic dispersion. We quantify this at the beginning of Sec.~\ref{sec:methods}. 
This property of multiple distance measures in the same galaxy is also the principle used to transfer distance measurements from other 
calibrators in building the local distance ladder. 
We use this fact to study the standardization procedure used in supernova cosmology using the \changes{SALT standardization relation}, as well as shed light on the 
physical processes underlying the color luminosity part of the relation.}\\

\changes{The  sibling pair under consideration has certain special properties that allow us to evade several systematics. These properties will not hold
 for siblings in general and so we discuss these.}

\changes{First, this sibling pair has an extremely small angular separation ($0.57 \arcsec\ $) corresponding to a projected distance of $0.6 \kpc\ .$ The 
angular separation is smaller than the scale at which the dust maps of the Milky Way appreciably vary, and thus the Milky way extinction, which is not perfectly
known affects them in an identical way, further reducing potential systematic uncertainties. 
An often discussed issue with such distance  measurements for \snia\ is the impact of the environment through local and global properties of the host galaxy. \revtwo{The local impacts are sometimes corrected for using host properties that are local to the line of sight. Despite the coincidental projected proximity, the pair of siblings under consideration may be still be physically affected by such environments due to a larger distance along the line of sight. However, the correction terms, which must be based on the projected local properties, must be zero in our particular example, as discussed in Sec.~\ref{sec:results} enabling us to emulate the results incorporating such corrections without determining the local host properties.} For more generic cases of \snia\ siblings, we are unlikely to repeat such proximity, and differences in \snia\ distances might need to take into account the differences in local stellar populations. While that would complicate the computation we did for other cases, this could also potentially help in resolving the difference between local and global dependence on galaxy properties.}\\

\changes{A second special property is the similarity of the light-curve shape parameter $x_1$ and difference in the color parameter $c$ of the sibling supernovae, which drives the strength of our constraints on $\beta$ and the relative insensitivity on $\alpha$. In general, for individual pairs of sibling supernovae, we will not get as tight constraints on $\beta$, and there might be higher sensitivity to the priors on $\alpha$. On the other hand, samples of sibling supernovae which would have a distribution of differences in parameters should enable using the same method to put tighter constraints on these parameters.}\\

\changes{Finally, we note that the two supernovae siblings may be found using different instruments (for e.g. in different surveys). In our case, we have been fortunate to obtain this sibling pair using the same instrument and thus our work is independent of inter-instrumental systematics}

\section{Conclusions}

Standardization of \snia\, and the characterization of systematic uncertainties in the process remains a crucial piece of supernova cosmology and its application to constraining the phenomenology of dark energy and the Hubble tension. Such systematics are particularly important when using supernova samples promised by ongoing and upcoming surveys where the large supernova sample size is expected to reduce the statistical uncertainties. Within the current industry standard of standardization based on the \salt\ model, an important aspect is the determination of population level parameter $\alpha, \beta$ that determine the linear importance of light curve shape and colour of \snia\, relative to their observed brightness in the standardization process.
In the conventional process, this determination of $\alpha, \beta$ can depend on the cosmological models used, the contribution due to environmental effects, and the Malmquist bias of the survey sample relative to the training sample. The estimates are often corrected for using catalogue simulations that forward model entire surveys and their selection function using input populations of \snia\ that are inferred from data. This is a complex process. Hence complementary and independent information determining these parameters would be extremely useful.\\

\changes{In this work, we use the fact that the difference in distances to sibling supernovae is negligible, to address standardization in a way which is independent of cosmology
 as well as several other systematics such as possible correlations of supernova brightnesses with local or global host galaxy properties. We introduced this method and applied 
 it to a pair of sibling} \snia\ (i.e. hosted in the same galaxy) obtained from the ZTF survey to constrain the parameter $\beta$ to $3.5 \pm 0.3.$ 
 While the historically reported values of $\beta$ have been much lower, their values have stabilized for the past few years. Our reported values 
 are consistent with the latest literature global values assuming a single $\beta$ for all SNe, but has a precision which is somewhat worse than 
 the current state-of-art. Its advantage is that it is independent of several of the systematic uncertainties of the current methodology. For example, 
 this does not depend on how accurately the population of \snia\ has been modeled in simulations, or how accurately the simulations can represent 
 complex time domain surveys. While we demonstrate this method with a particular pair of siblings, where the constraints on $\beta$ was expected to be strong (due to the differences in the best fit values of the \snia\ model parameters), the method can be extended to samples of siblings.  This would not have been easy in the past, as sibling \snia\ are rare. Thanks to the dramatic improvement in sky survey volume, ongoing surveys are reporting several \snia\ siblings. As we continue to wide-field surveys like ZTF Phase II, Young Supernova Experiment (YSE) and the Large Synoptic Survey Telescope (LSST)  on the Vera C. Rubin Observatory, sibling \snia\ will be much more commonly found, allowing this method to be applied to a sample for accurate tests of the \snia] standardization needed for precision cosmology. If the sample contains siblings whose light curve shape parameters differ significantly as well, this method can constrain both $\alpha,$ and $\beta$ without other priors. Significant work has gone into developing standardization models~\citep[see e.g.][]{2018ApJ...869..167S,2020A&A...636A..46L,2019ApJ...871..219H}, that are expected to improve upon {\salt}. Such models involve different (and more) free parameters that need to be determined in a manner similar $\alpha$ and $\beta$ in the \salt\ model. Presumably, this method can be applied in a similar way, though we have not tried this at all.\\

A different path to better standardization is improving our understanding of the physical processes underlying the success. One of the most promising routes has been interpreting the physics underlying the colour dependent term standardization. This could be attributed to intrinsic colour diversity of \snia\ and physics connecting this to the brightness, or/and extinction due to dust in the interstellar medium of the host galaxy, and possibly in the circumstellar environment. Studies for this have used statistical sub-samples of \snia\ based on hosts, or samples of individual \snia\ where the extinction parameters are determined through measurements at multiple wavelengths, including space observations in the near-UV  \citep[see e.g.][]{2015MNRAS.453.3300A}, or even into the mid-IR \citep{2017MNRAS.466.3442J}. Again using the independence of distance/brightness differences of \snia\ in the same galaxy, sibling \snia\ promise an alternative way of probing extinction through accurate measurements of attenuation with well-calibrated systems.\\

\section{Acknowledgements}

RB was supported by the research project grant "Understanding the Dynamic Universe" funded by the Knut and Alice Wallenberg Foundation under Dnr KAW 2018.0067. AG acknowledges support from the Swedish Research Council under Dnr VR 2016-03274 and 2020-03444. SD acknowledges support from the Isaac Newton Trust and the Kavli Foundation through Newton-Kavli fellowship. MR, MS and YLK have received funding from the European Research Council (ERC) under the European Union's Horizon 2020 research and innovation programme (grant agreement n°759194 - USNAC). ECK acknowledges support from the G.R.E.A.T research environment funded by {\em Vetenskapsr\aa det}, the Swedish Research Council, under project number 2016-06012, and support from The Wenner-Gren Foundations.

Based on observations obtained with the Samuel Oschin Telescope 48-inch and the 60-inch Telescope at the Palomar Observatory as part of the Zwicky Transient Facility project. ZTF is supported by the National Science Foundation under Grant No. AST-1440341 and a collaboration including Caltech, IPAC, the Weizmann Institute for Science, the Oskar Klein Center at Stockholm University, the University of Maryland, the University of Washington, Deutsches Elektronen-Synchrotron and Humboldt University, Los Alamos National Laboratories, the TANGO Consortium of Taiwan, the University of Wisconsin at Milwaukee, and Lawrence Berkeley National Laboratories. Operations are conducted by COO, IPAC, and UW.
The ZTF forced-photometry service was funded under the Heising-Simons Foundation grant. This work was supported by the GROWTH project funded by the National Science Foundation under Grant No 1545949. 

Based on observations made with the Nordic Optical Telescope, owned in collaboration by the University of Turku and Aarhus University, and operated jointly by Aarhus University, the University of Turku and the University of Oslo, representing Denmark, Finland and Norway, the University of Iceland and Stockholm University at the Observatorio del Roque de los Muchachos, La Palma, Spain, of the Instituto de Astrofisica de Canarias.


\subsection{Data Availability} The ZTF images used for forced photometry are available at IRSA \url{https://irsa.ipac.caltech.edu/Missions/ztf.html} and through the ZTF data portal~\url{https://www.ztf.caltech.edu/page/dr5#12}. The alerts used for initial processing are made available by ZTF in the form of an alert archive~\url{https://ztf.uw.edu/}, as well as through public brokers that access that ZTF alerts in near real time. The results of running the forced photometry pipelines which form the main data used in the analysis of this article are provided in the form of tables in appendix C. The software used for the analysis and is described below.

software: {Numpy~\citep{2011CSE....13b..22V}, Astropy~\citep{2013A&A...558A..33A,2018AJ....156..123A},~sncosmo~\citep{2016ascl.soft11017B}, PhotUtils~\citep{Bradley_2019_2533376}, ZTFQuery~\citep{2018zndo...1345222R}, SWARP~\citep{2010ascl.soft10068B}, HOTPANTS~\citep{2015ascl.soft04004B}, ZUDS, fringez~\citep{2021arXiv210210738M}, LAMBDAR~\cite{Wright2016a}, Prospector~\citep{Leja2017a}, emcee~\citep{2013PASP..125..306F}, pygtc~\citep{2016JOSS....1...46B}, matplotlib~\citep{2007CSE.....9...90H}, IPAC forced Photometry Service.}\\



\bibliographystyle{mnras}
\bibliography{bib/paper} 

\begin{thebibliography}{}
\makeatletter
\relax
\def\mn@urlcharsother{\let\do\@makeother \do\$\do\&\do\#\do\^\do\_\do\%\do\~}
\def\mn@doi{\begingroup\mn@urlcharsother \@ifnextchar [ {\mn@doi@}
  {\mn@doi@[]}}
\def\mn@doi@[#1]#2{\def\@tempa{#1}\ifx\@tempa\@empty \href
  {http://dx.doi.org/#2} {doi:#2}\else \href {http://dx.doi.org/#2} {#1}\fi
  \endgroup}
\def\mn@eprint#1#2{\mn@eprint@#1:#2::\@nil}
\def\mn@eprint@arXiv#1{\href {http://arxiv.org/abs/#1} {{\tt arXiv:#1}}}
\def\mn@eprint@dblp#1{\href {http://dblp.uni-trier.de/rec/bibtex/#1.xml}
  {dblp:#1}}
\def\mn@eprint@#1:#2:#3:#4\@nil{\def\@tempa {#1}\def\@tempb {#2}\def\@tempc
  {#3}\ifx \@tempc \@empty \let \@tempc \@tempb \let \@tempb \@tempa \fi \ifx
  \@tempb \@empty \def\@tempb {arXiv}\fi \@ifundefined
  {mn@eprint@\@tempb}{\@tempb:\@tempc}{\expandafter \expandafter \csname
  mn@eprint@\@tempb\endcsname \expandafter{\@tempc}}}

\bibitem[\protect\citeauthoryear{{Ahn} et~al.,}{{Ahn} et~al.}{2012}]{Ahn2012a}
{Ahn} C.~P.,  et~al., 2012, \mn@doi [\apjs] {10.1088/0067-0049/203/2/21}, \href
  {https://ui.adsabs.harvard.edu/abs/2012ApJS..203...21A} {203, 21}

\bibitem[\protect\citeauthoryear{{Alard} \& {Lupton}}{{Alard} \&
  {Lupton}}{1998}]{1998ApJ...503..325A}
{Alard} C.,  {Lupton} R.~H.,  1998, \mn@doi [\apj] {10.1086/305984}, \href
  {https://ui.adsabs.harvard.edu/abs/1998ApJ...503..325A} {503, 325}

\bibitem[\protect\citeauthoryear{{Amanullah} et~al.,}{{Amanullah}
  et~al.}{2010}]{2010ApJ...716..712A}
{Amanullah} R.,  et~al., 2010, \mn@doi [\apj] {10.1088/0004-637X/716/1/712},
  \href {https://ui.adsabs.harvard.edu/abs/2010ApJ...716..712A} {716, 712}

\bibitem[\protect\citeauthoryear{{Amanullah} et~al.,}{{Amanullah}
  et~al.}{2014}]{2014ApJ...788L..21A}
{Amanullah} R.,  et~al., 2014, \mn@doi [\apjl] {10.1088/2041-8205/788/2/L21},
  \href {https://ui.adsabs.harvard.edu/abs/2014ApJ...788L..21A} {788, L21}

\bibitem[\protect\citeauthoryear{{Amanullah} et~al.,}{{Amanullah}
  et~al.}{2015}]{2015MNRAS.453.3300A}
{Amanullah} R.,  et~al., 2015, \mn@doi [\mnras] {10.1093/mnras/stv1505}, \href
  {https://ui.adsabs.harvard.edu/abs/2015MNRAS.453.3300A} {453, 3300}

\bibitem[\protect\citeauthoryear{{Astier} et~al.,}{{Astier}
  et~al.}{2006}]{2006A&A...447...31A}
{Astier} P.,  et~al., 2006, \mn@doi [\aap] {10.1051/0004-6361:20054185}, \href
  {https://ui.adsabs.harvard.edu/abs/2006A&A...447...31A} {447, 31}

\bibitem[\protect\citeauthoryear{{Astropy Collaboration} et~al.,}{{Astropy
  Collaboration} et~al.}{2013}]{2013A&A...558A..33A}
{Astropy Collaboration} et~al., 2013, \mn@doi [\aap]
  {10.1051/0004-6361/201322068}, \href
  {https://ui.adsabs.harvard.edu/abs/2013A&A...558A..33A} {558, A33}

\bibitem[\protect\citeauthoryear{{Astropy Collaboration} et~al.,}{{Astropy
  Collaboration} et~al.}{2018}]{2018AJ....156..123A}
{Astropy Collaboration} et~al., 2018, \mn@doi [\aj] {10.3847/1538-3881/aabc4f},
  \href {https://ui.adsabs.harvard.edu/abs/2018AJ....156..123A} {156, 123}

\bibitem[\protect\citeauthoryear{{Barbary} et~al.,}{{Barbary}
  et~al.}{2016}]{2016ascl.soft11017B}
{Barbary} K.,  et~al., 2016, {SNCosmo: Python library for supernova cosmology}
  (\mn@eprint {ascl} {1611.017})

\bibitem[\protect\citeauthoryear{{Becker}}{{Becker}}{2015}]{2015ascl.soft04004B}
{Becker} A.,  2015, {HOTPANTS: High Order Transform of PSF ANd Template
  Subtraction} (\mn@eprint {ascl} {1504.004})

\bibitem[\protect\citeauthoryear{{Bellm} et~al.,}{{Bellm}
  et~al.}{2019a}]{2019PASP..131a8002B}
{Bellm} E.~C.,  et~al., 2019a, \mn@doi [\pasp] {10.1088/1538-3873/aaecbe},
  \href {https://ui.adsabs.harvard.edu/abs/2019PASP..131a8002B} {131, 018002}

\bibitem[\protect\citeauthoryear{{Bellm} et~al.,}{{Bellm}
  et~al.}{2019b}]{2019PASP..131f8003B}
{Bellm} E.~C.,  et~al., 2019b, \mn@doi [\pasp] {10.1088/1538-3873/ab0c2a},
  \href {https://ui.adsabs.harvard.edu/abs/2019PASP..131f8003B} {131, 068003}

\bibitem[\protect\citeauthoryear{{Bertin}}{{Bertin}}{2010}]{2010ascl.soft10068B}
{Bertin} E.,  2010, {SWarp: Resampling and Co-adding FITS Images Together}
  (\mn@eprint {ascl} {1010.068})

\bibitem[\protect\citeauthoryear{{Betoule} et~al.,}{{Betoule}
  et~al.}{2014}]{2014A&A...568A..22B}
{Betoule} M.,  et~al., 2014, \mn@doi [\aap] {10.1051/0004-6361/201423413},
  \href {https://ui.adsabs.harvard.edu/abs/2014A&A...568A..22B} {568, A22}

\bibitem[\protect\citeauthoryear{{Bocquet} \& {Carter}}{{Bocquet} \&
  {Carter}}{2016}]{2016JOSS....1...46B}
{Bocquet} S.,  {Carter} F.~W.,  2016, \mn@doi [The Journal of Open Source
  Software] {10.21105/joss.00046}, \href
  {https://ui.adsabs.harvard.edu/abs/2016JOSS....1...46B} {1, 46}

\bibitem[\protect\citeauthoryear{{Bourne} et~al.,}{{Bourne}
  et~al.}{2012}]{Bourne2012a}
{Bourne} N.,  et~al., 2012, \mn@doi [\mnras]
  {10.1111/j.1365-2966.2012.20528.x}, \href
  {https://ui.adsabs.harvard.edu/abs/2012MNRAS.421.3027B} {421, 3027}

\bibitem[\protect\citeauthoryear{{Bradley} et~al.,}{{Bradley}
  et~al.}{2019}]{Bradley_2019_2533376}
{Bradley} L.,  et~al., 2019, {astropy/photutils: v0.6},
  \mn@doi{10.5281/zenodo.2533376}

\bibitem[\protect\citeauthoryear{{Brout} \& {Scolnic}}{{Brout} \&
  {Scolnic}}{2021}]{2020arXiv200410206B}
{Brout} D.,  {Scolnic} D.,  2021, \mn@doi [\apj] {10.3847/1538-4357/abd69b},
  \href {https://ui.adsabs.harvard.edu/abs/2021ApJ...909...26B} {909, 26}

\bibitem[\protect\citeauthoryear{{Brout} et~al.,}{{Brout}
  et~al.}{2019}]{2019ApJ...874..150B}
{Brout} D.,  et~al., 2019, \mn@doi [\apj] {10.3847/1538-4357/ab08a0}, \href
  {https://ui.adsabs.harvard.edu/abs/2019ApJ...874..150B} {874, 150}

\bibitem[\protect\citeauthoryear{{Bulla}, {Goobar}  \& {Dhawan}}{{Bulla}
  et~al.}{2018}]{2018MNRAS.479.3663B}
{Bulla} M.,  {Goobar} A.,   {Dhawan} S.,  2018, \mn@doi [\mnras]
  {10.1093/mnras/sty1619}, \href
  {https://ui.adsabs.harvard.edu/abs/2018MNRAS.479.3663B} {479, 3663}

\bibitem[\protect\citeauthoryear{{Burns} et~al.,}{{Burns}
  et~al.}{2011}]{2011AJ....141...19B}
{Burns} C.~R.,  et~al., 2011, \mn@doi [\aj] {10.1088/0004-6256/141/1/19}, \href
  {https://ui.adsabs.harvard.edu/abs/2011AJ....141...19B} {141, 19}

\bibitem[\protect\citeauthoryear{{Burns} et~al.,}{{Burns}
  et~al.}{2014}]{2014ApJ...789...32B}
{Burns} C.~R.,  et~al., 2014, \mn@doi [\apj] {10.1088/0004-637X/789/1/32},
  \href {https://ui.adsabs.harvard.edu/abs/2014ApJ...789...32B} {789, 32}

\bibitem[\protect\citeauthoryear{{Burns} et~al.,}{{Burns}
  et~al.}{2020}]{2020ApJ...895..118B}
{Burns} C.~R.,  et~al., 2020, \mn@doi [\apj] {10.3847/1538-4357/ab8e3e}, \href
  {https://ui.adsabs.harvard.edu/abs/2020ApJ...895..118B} {895, 118}

\bibitem[\protect\citeauthoryear{{Byler}, {Dalcanton}, {Conroy}  \&
  {Johnson}}{{Byler} et~al.}{2017}]{Byler2017a}
{Byler} N.,  {Dalcanton} J.~J.,  {Conroy} C.,   {Johnson} B.~D.,  2017, \mn@doi
  [\apj] {10.3847/1538-4357/aa6c66}, \href
  {https://ui.adsabs.harvard.edu/abs/2017ApJ...840...44B} {840, 44}

\bibitem[\protect\citeauthoryear{{Calzetti}, {Armus}, {Bohlin}, {Kinney},
  {Koornneef}  \& {Storchi-Bergmann}}{{Calzetti} et~al.}{2000}]{Calzetti2000a}
{Calzetti} D.,  {Armus} L.,  {Bohlin} R.~C.,  {Kinney} A.~L.,  {Koornneef} J.,
   {Storchi-Bergmann} T.,  2000, \mn@doi [\apj] {10.1086/308692}, \href
  {https://ui.adsabs.harvard.edu/abs/2000ApJ...533..682C} {533, 682}

\bibitem[\protect\citeauthoryear{{Cardelli}, {Clayton}  \& {Mathis}}{{Cardelli}
  et~al.}{1989}]{1989ApJ...345..245C}
{Cardelli} J.~A.,  {Clayton} G.~C.,   {Mathis} J.~S.,  1989, \mn@doi [\apj]
  {10.1086/167900}, \href
  {https://ui.adsabs.harvard.edu/abs/1989ApJ...345..245C} {345, 245}

\bibitem[\protect\citeauthoryear{{Chabrier}}{{Chabrier}}{2003}]{Chabrier2003a}
{Chabrier} G.,  2003, \mn@doi [\pasp] {10.1086/376392}, \href
  {https://ui.adsabs.harvard.edu/abs/2003PASP..115..763C} {115, 763}

\bibitem[\protect\citeauthoryear{{Chambers} et~al.,}{{Chambers}
  et~al.}{2016}]{Chambers2016a}
{Chambers} K.~C.,  et~al., 2016, arXiv e-prints, \href
  {https://ui.adsabs.harvard.edu/abs/2016arXiv161205560C} {p. arXiv:1612.05560}

\bibitem[\protect\citeauthoryear{{Childress} et~al.,}{{Childress}
  et~al.}{2013}]{2013ApJ...770..108C}
{Childress} M.,  et~al., 2013, \mn@doi [\apj] {10.1088/0004-637X/770/2/108},
  \href {https://ui.adsabs.harvard.edu/abs/2013ApJ...770..108C} {770, 108}

\bibitem[\protect\citeauthoryear{{Chotard} et~al.,}{{Chotard}
  et~al.}{2011}]{2011A&A...529L...4C}
{Chotard} N.,  et~al., 2011, \mn@doi [\aap] {10.1051/0004-6361/201116723},
  \href {https://ui.adsabs.harvard.edu/abs/2011A&A...529L...4C} {529, L4}

\bibitem[\protect\citeauthoryear{{Dettman} et~al.,}{{Dettman}
  et~al.}{2021}]{2021arXiv210206524D}
{Dettman} K.~G.,  et~al., 2021, arXiv e-prints, \href
  {https://ui.adsabs.harvard.edu/abs/2021arXiv210206524D} {p. arXiv:2102.06524}

\bibitem[\protect\citeauthoryear{{Foley} et~al.,}{{Foley}
  et~al.}{2014}]{2014MNRAS.443.2887F}
{Foley} R.~J.,  et~al., 2014, \mn@doi [\mnras] {10.1093/mnras/stu1378}, \href
  {https://ui.adsabs.harvard.edu/abs/2014MNRAS.443.2887F} {443, 2887}

\bibitem[\protect\citeauthoryear{{Foley} et~al.,}{{Foley}
  et~al.}{2018}]{2018MNRAS.475..193F}
{Foley} R.~J.,  et~al., 2018, \mn@doi [\mnras] {10.1093/mnras/stx3136}, \href
  {https://ui.adsabs.harvard.edu/abs/2018MNRAS.475..193F} {475, 193}

\bibitem[\protect\citeauthoryear{{Foreman-Mackey}, {Hogg}, {Lang}  \&
  {Goodman}}{{Foreman-Mackey} et~al.}{2013}]{2013PASP..125..306F}
{Foreman-Mackey} D.,  {Hogg} D.~W.,  {Lang} D.,   {Goodman} J.,  2013, \mn@doi
  [\pasp] {10.1086/670067}, \href
  {https://ui.adsabs.harvard.edu/abs/2013PASP..125..306F} {125, 306}

\bibitem[\protect\citeauthoryear{{Fremling} et~al.,}{{Fremling}
  et~al.}{2020}]{2020ApJ...895...32F}
{Fremling} C.,  et~al., 2020, \mn@doi [\apj] {10.3847/1538-4357/ab8943}, \href
  {https://ui.adsabs.harvard.edu/abs/2020ApJ...895...32F} {895, 32}

\bibitem[\protect\citeauthoryear{{Gagliano}, {Narayan}, {Engel}, {Carrasco
  Kind}  \& {LSST Dark Energy Science Collaboration}}{{Gagliano}
  et~al.}{2021}]{2020arXiv200809630G}
{Gagliano} A.,  {Narayan} G.,  {Engel} A.,  {Carrasco Kind} M.,   {LSST Dark
  Energy Science Collaboration} 2021, \mn@doi [\apj]
  {10.3847/1538-4357/abd02b}, \href
  {https://ui.adsabs.harvard.edu/abs/2021ApJ...908..170G} {908, 170}

\bibitem[\protect\citeauthoryear{{Galbany} et~al.,}{{Galbany}
  et~al.}{2012}]{2012ApJ...755..125G}
{Galbany} L.,  et~al., 2012, \mn@doi [\apj] {10.1088/0004-637X/755/2/125},
  \href {https://ui.adsabs.harvard.edu/abs/2012ApJ...755..125G} {755, 125}

\bibitem[\protect\citeauthoryear{{Goobar}}{{Goobar}}{2008}]{2008ApJ...686L.103G}
{Goobar} A.,  2008, \mn@doi [\apjl] {10.1086/593060}, \href
  {https://ui.adsabs.harvard.edu/abs/2008ApJ...686L.103G} {686, L103}

\bibitem[\protect\citeauthoryear{{Goobar} \& {Leibundgut}}{{Goobar} \&
  {Leibundgut}}{2011}]{2011ARNPS..61..251G}
{Goobar} A.,  {Leibundgut} B.,  2011, \mn@doi [Annual Review of Nuclear and
  Particle Science] {10.1146/annurev-nucl-102010-130434}, \href
  {https://ui.adsabs.harvard.edu/abs/2011ARNPS..61..251G} {61, 251}

\bibitem[\protect\citeauthoryear{{Goobar} et~al.,}{{Goobar}
  et~al.}{2014}]{2014ApJ...784L..12G}
{Goobar} A.,  et~al., 2014, \mn@doi [\apjl] {10.1088/2041-8205/784/1/L12},
  \href {https://ui.adsabs.harvard.edu/abs/2014ApJ...784L..12G} {784, L12}

\bibitem[\protect\citeauthoryear{{Graham} et~al.,}{{Graham}
  et~al.}{2019}]{2019PASP..131g8001G}
{Graham} M.~J.,  et~al., 2019, \mn@doi [\pasp] {10.1088/1538-3873/ab006c},
  \href {https://ui.adsabs.harvard.edu/abs/2019PASP..131g8001G} {131, 078001}

\bibitem[\protect\citeauthoryear{{Guy}, {Astier}, {Nobili}, {Regnault}  \&
  {Pain}}{{Guy} et~al.}{2005}]{2005A&A...443..781G}
{Guy} J.,  {Astier} P.,  {Nobili} S.,  {Regnault} N.,   {Pain} R.,  2005,
  \mn@doi [\aap] {10.1051/0004-6361:20053025}, \href
  {https://ui.adsabs.harvard.edu/abs/2005A&A...443..781G} {443, 781}

\bibitem[\protect\citeauthoryear{{Guy} et~al.,}{{Guy}
  et~al.}{2007}]{2007A&A...466...11G}
{Guy} J.,  et~al., 2007, \mn@doi [\aap] {10.1051/0004-6361:20066930}, \href
  {https://ui.adsabs.harvard.edu/abs/2007A&A...466...11G} {466, 11}

\bibitem[\protect\citeauthoryear{{Guy} et~al.,}{{Guy}
  et~al.}{2010}]{2010A&A...523A...7G}
{Guy} J.,  et~al., 2010, \mn@doi [\aap] {10.1051/0004-6361/201014468}, \href
  {https://ui.adsabs.harvard.edu/abs/2010A&A...523A...7G} {523, A7}

\bibitem[\protect\citeauthoryear{{Hayden}, {Rubin}  \& {Strovink}}{{Hayden}
  et~al.}{2019}]{2019ApJ...871..219H}
{Hayden} B.,  {Rubin} D.,   {Strovink} M.,  2019, \mn@doi [\apj]
  {10.3847/1538-4357/aaf232}, \href
  {https://ui.adsabs.harvard.edu/abs/2019ApJ...871..219H} {871, 219}

\bibitem[\protect\citeauthoryear{{Hinton} et~al.,}{{Hinton}
  et~al.}{2019}]{2019ApJ...876...15H}
{Hinton} S.~R.,  et~al., 2019, \mn@doi [\apj] {10.3847/1538-4357/ab13a3}, \href
  {https://ui.adsabs.harvard.edu/abs/2019ApJ...876...15H} {876, 15}

\bibitem[\protect\citeauthoryear{{Hoang}}{{Hoang}}{2017}]{2017ApJ...836...13H}
{Hoang} T.,  2017, \mn@doi [\apj] {10.3847/1538-4357/836/1/13}, \href
  {https://ui.adsabs.harvard.edu/abs/2017ApJ...836...13H} {836, 13}

\bibitem[\protect\citeauthoryear{{Hoang}}{{Hoang}}{2021}]{2021ApJ...907...37H}
{Hoang} T.,  2021, \mn@doi [\apj] {10.3847/1538-4357/abccc8}, \href
  {https://ui.adsabs.harvard.edu/abs/2021ApJ...907...37H} {907, 37}

\bibitem[\protect\citeauthoryear{{Hunter}}{{Hunter}}{2007}]{2007CSE.....9...90H}
{Hunter} J.~D.,  2007, \mn@doi [Computing in Science and Engineering]
  {10.1109/MCSE.2007.55}, \href
  {https://ui.adsabs.harvard.edu/abs/2007CSE.....9...90H} {9, 90}

\bibitem[\protect\citeauthoryear{{Ivezi{\'c}} et~al.,}{{Ivezi{\'c}}
  et~al.}{2019}]{2019ApJ...873..111I}
{Ivezi{\'c}} {\v{Z}}.,  et~al., 2019, \mn@doi [\apj]
  {10.3847/1538-4357/ab042c}, \href
  {https://ui.adsabs.harvard.edu/abs/2019ApJ...873..111I} {873, 111}

\bibitem[\protect\citeauthoryear{{Johansson} et~al.,}{{Johansson}
  et~al.}{2017}]{2017MNRAS.466.3442J}
{Johansson} J.,  et~al., 2017, \mn@doi [\mnras] {10.1093/mnras/stw3350}, \href
  {https://ui.adsabs.harvard.edu/abs/2017MNRAS.466.3442J} {466, 3442}

\bibitem[\protect\citeauthoryear{{Johansson} et~al.,}{{Johansson}
  et~al.}{2021}]{2021arXiv210506236J}
{Johansson} J.,  et~al., 2021, arXiv e-prints, \href
  {https://ui.adsabs.harvard.edu/abs/2021arXiv210506236J} {p. arXiv:2105.06236}

\bibitem[\protect\citeauthoryear{{Jones} et~al.,}{{Jones}
  et~al.}{2018a}]{2018ApJ...857...51J}
{Jones} D.~O.,  et~al., 2018a, \mn@doi [\apj] {10.3847/1538-4357/aab6b1}, \href
  {https://ui.adsabs.harvard.edu/abs/2018ApJ...857...51J} {857, 51}

\bibitem[\protect\citeauthoryear{{Jones} et~al.,}{{Jones}
  et~al.}{2018b}]{2018ApJ...867..108J}
{Jones} D.~O.,  et~al., 2018b, \mn@doi [\apj] {10.3847/1538-4357/aae2b9}, \href
  {https://ui.adsabs.harvard.edu/abs/2018ApJ...867..108J} {867, 108}

\bibitem[\protect\citeauthoryear{{Jones} et~al.,}{{Jones}
  et~al.}{2021}]{2020arXiv201009724J}
{Jones} D.~O.,  et~al., 2021, \mn@doi [\apj] {10.3847/1538-4357/abd7f5}, \href
  {https://ui.adsabs.harvard.edu/abs/2021ApJ...908..143J} {908, 143}

\bibitem[\protect\citeauthoryear{{Kasliwal} et~al.,}{{Kasliwal}
  et~al.}{2019}]{Kasliwal_2019}
{Kasliwal} M.~M.,  et~al., 2019, \mn@doi [\pasp] {10.1088/1538-3873/aafbc2},
  \href {https://ui.adsabs.harvard.edu/abs/2019PASP..131c8003K} {131, 038003}

\bibitem[\protect\citeauthoryear{{Kelly}, {Hicken}, {Burke}, {Mandel}  \&
  {Kirshner}}{{Kelly} et~al.}{2010}]{2010ApJ...715..743K}
{Kelly} P.~L.,  {Hicken} M.,  {Burke} D.~L.,  {Mandel} K.~S.,   {Kirshner}
  R.~P.,  2010, \mn@doi [\apj] {10.1088/0004-637X/715/2/743}, \href
  {https://ui.adsabs.harvard.edu/abs/2010ApJ...715..743K} {715, 743}

\bibitem[\protect\citeauthoryear{{Kelsey} et~al.,}{{Kelsey}
  et~al.}{2021}]{2021MNRAS.501.4861K}
{Kelsey} L.,  et~al., 2021, \mn@doi [\mnras] {10.1093/mnras/staa3924}, \href
  {https://ui.adsabs.harvard.edu/abs/2021MNRAS.501.4861K} {501, 4861}

\bibitem[\protect\citeauthoryear{{Kessler} \& {Scolnic}}{{Kessler} \&
  {Scolnic}}{2017}]{2017ApJ...836...56K}
{Kessler} R.,  {Scolnic} D.,  2017, \mn@doi [\apj]
  {10.3847/1538-4357/836/1/56}, \href
  {https://ui.adsabs.harvard.edu/abs/2017ApJ...836...56K} {836, 56}

\bibitem[\protect\citeauthoryear{{Kessler} et~al.,}{{Kessler}
  et~al.}{2009}]{2009ApJS..185...32K}
{Kessler} R.,  et~al., 2009, \mn@doi [\apjs] {10.1088/0067-0049/185/1/32},
  \href {https://ui.adsabs.harvard.edu/abs/2009ApJS..185...32K} {185, 32}

\bibitem[\protect\citeauthoryear{{Kessler} et~al.,}{{Kessler}
  et~al.}{2013}]{2013ApJ...764...48K}
{Kessler} R.,  et~al., 2013, \mn@doi [\apj] {10.1088/0004-637X/764/1/48}, \href
  {https://ui.adsabs.harvard.edu/abs/2013ApJ...764...48K} {764, 48}

\bibitem[\protect\citeauthoryear{{Kessler} et~al.,}{{Kessler}
  et~al.}{2019}]{2019MNRAS.485.1171K}
{Kessler} R.,  et~al., 2019, \mn@doi [\mnras] {10.1093/mnras/stz463}, \href
  {https://ui.adsabs.harvard.edu/abs/2019MNRAS.485.1171K} {485, 1171}

\bibitem[\protect\citeauthoryear{{LSST Science Collaboration} et~al.,}{{LSST
  Science Collaboration} et~al.}{2009}]{2009arXiv0912.0201L}
{LSST Science Collaboration} et~al., 2009, arXiv e-prints, \href
  {https://ui.adsabs.harvard.edu/abs/2009arXiv0912.0201L} {p. arXiv:0912.0201}

\bibitem[\protect\citeauthoryear{{Lampeitl} et~al.,}{{Lampeitl}
  et~al.}{2010}]{2010ApJ...722..566L}
{Lampeitl} H.,  et~al., 2010, \mn@doi [\apj] {10.1088/0004-637X/722/1/566},
  \href {https://ui.adsabs.harvard.edu/abs/2010ApJ...722..566L} {722, 566}

\bibitem[\protect\citeauthoryear{{Lang}}{{Lang}}{2014}]{Lang2014a}
{Lang} D.,  2014, \mn@doi [\aj] {10.1088/0004-6256/147/5/108}, \href
  {https://ui.adsabs.harvard.edu/abs/2014AJ....147..108L} {147, 108}

\bibitem[\protect\citeauthoryear{{L{\'e}get} et~al.,}{{L{\'e}get}
  et~al.}{2020}]{2020A&A...636A..46L}
{L{\'e}get} P.~F.,  et~al., 2020, \mn@doi [\aap] {10.1051/0004-6361/201834954},
  \href {https://ui.adsabs.harvard.edu/abs/2020A&A...636A..46L} {636, A46}

\bibitem[\protect\citeauthoryear{{Leja}, {Johnson}, {Conroy}, {van Dokkum}  \&
  {Byler}}{{Leja} et~al.}{2017}]{Leja2017a}
{Leja} J.,  {Johnson} B.~D.,  {Conroy} C.,  {van Dokkum} P.~G.,   {Byler} N.,
  2017, \mn@doi [\apj] {10.3847/1538-4357/aa5ffe}, \href
  {https://ui.adsabs.harvard.edu/abs/2017ApJ...837..170L} {837, 170}

\bibitem[\protect\citeauthoryear{{Licquia} \& {Newman}}{{Licquia} \&
  {Newman}}{2015}]{2015ApJ...806...96L}
{Licquia} T.~C.,  {Newman} J.~A.,  2015, \mn@doi [\apj]
  {10.1088/0004-637X/806/1/96}, \href
  {https://ui.adsabs.harvard.edu/abs/2015ApJ...806...96L} {806, 96}

\bibitem[\protect\citeauthoryear{{Liddle}}{{Liddle}}{2007}]{2007MNRAS.377L..74L}
{Liddle} A.~R.,  2007, \mn@doi [\mnras] {10.1111/j.1745-3933.2007.00306.x},
  \href {https://ui.adsabs.harvard.edu/abs/2007MNRAS.377L..74L} {377, L74}

\bibitem[\protect\citeauthoryear{{Maeda}, {Nozawa}, {Nagao}  \&
  {Motohara}}{{Maeda} et~al.}{2015}]{2015MNRAS.452.3281M}
{Maeda} K.,  {Nozawa} T.,  {Nagao} T.,   {Motohara} K.,  2015, \mn@doi [\mnras]
  {10.1093/mnras/stv1498}, \href
  {https://ui.adsabs.harvard.edu/abs/2015MNRAS.452.3281M} {452, 3281}

\bibitem[\protect\citeauthoryear{{Mainzer} et~al.,}{{Mainzer}
  et~al.}{2014}]{Mainzer2014a}
{Mainzer} A.,  et~al., 2014, \mn@doi [\apj] {10.1088/0004-637X/792/1/30}, \href
  {https://ui.adsabs.harvard.edu/abs/2014ApJ...792...30M} {792, 30}

\bibitem[\protect\citeauthoryear{{Marriner} et~al.,}{{Marriner}
  et~al.}{2011}]{2011ApJ...740...72M}
{Marriner} J.,  et~al., 2011, \mn@doi [\apj] {10.1088/0004-637X/740/2/72},
  \href {https://ui.adsabs.harvard.edu/abs/2011ApJ...740...72M} {740, 72}

\bibitem[\protect\citeauthoryear{{Masci} et~al.,}{{Masci}
  et~al.}{2019}]{2019PASP..131a8003M}
{Masci} F.~J.,  et~al., 2019, \mn@doi [\pasp] {10.1088/1538-3873/aae8ac}, \href
  {https://ui.adsabs.harvard.edu/abs/2019PASP..131a8003M} {131, 018003}

\bibitem[\protect\citeauthoryear{{Medford} et~al.,}{{Medford}
  et~al.}{2021}]{2021arXiv210210738M}
{Medford} M.~S.,  et~al., 2021, \mn@doi [\pasp] {10.1088/1538-3873/abfe9d},
  \href {https://ui.adsabs.harvard.edu/abs/2021PASP..133f4503M} {133, 064503}

\bibitem[\protect\citeauthoryear{{Meisner}, {Lang}  \& {Schlegel}}{{Meisner}
  et~al.}{2017}]{Meisner2017a}
{Meisner} A.~M.,  {Lang} D.,   {Schlegel} D.~J.,  2017, \mn@doi [\aj]
  {10.3847/1538-3881/153/1/38}, \href
  {https://ui.adsabs.harvard.edu/abs/2017AJ....153...38M} {153, 38}

\bibitem[\protect\citeauthoryear{{Mosher} et~al.,}{{Mosher}
  et~al.}{2014}]{2014ApJ...793...16M}
{Mosher} J.,  et~al., 2014, \mn@doi [\apj] {10.1088/0004-637X/793/1/16}, \href
  {https://ui.adsabs.harvard.edu/abs/2014ApJ...793...16M} {793, 16}

\bibitem[\protect\citeauthoryear{{Nataf} et~al.,}{{Nataf}
  et~al.}{2016}]{2016MNRAS.456.2692N}
{Nataf} D.~M.,  et~al., 2016, \mn@doi [\mnras] {10.1093/mnras/stv2843}, \href
  {https://ui.adsabs.harvard.edu/abs/2016MNRAS.456.2692N} {456, 2692}

\bibitem[\protect\citeauthoryear{{Nobili} \& {Goobar}}{{Nobili} \&
  {Goobar}}{2008}]{2008A&A...487...19N}
{Nobili} S.,  {Goobar} A.,  2008, \mn@doi [\aap] {10.1051/0004-6361:20079292},
  \href {https://ui.adsabs.harvard.edu/abs/2008A&A...487...19N} {487, 19}

\bibitem[\protect\citeauthoryear{{Nordin}, {Brinnel}, {Giomi}, {Santen},
  {Gal-Yam}, {Yaron}  \& {Schulze}}{{Nordin}
  et~al.}{2019}]{2019TNSTR1202....1N}
{Nordin} J.,  {Brinnel} V.,  {Giomi} M.,  {Santen} J.~V.,  {Gal-Yam} A.,
  {Yaron} O.,   {Schulze} S.,  2019, Transient Name Server Discovery Report,
  \href {https://ui.adsabs.harvard.edu/abs/2019TNSTR1202....1N} {2019-1202, 1}

\bibitem[\protect\citeauthoryear{{Nugent}, {Kim}  \& {Perlmutter}}{{Nugent}
  et~al.}{2002}]{2002PASP..114..803N}
{Nugent} P.,  {Kim} A.,   {Perlmutter} S.,  2002, \mn@doi [\pasp]
  {10.1086/341707}, \href
  {https://ui.adsabs.harvard.edu/abs/2002PASP..114..803N} {114, 803}

\bibitem[\protect\citeauthoryear{{Perley} et~al.,}{{Perley}
  et~al.}{2020}]{2020ApJ...904...35P}
{Perley} D.~A.,  et~al., 2020, \mn@doi [\apj] {10.3847/1538-4357/abbd98}, \href
  {https://ui.adsabs.harvard.edu/abs/2020ApJ...904...35P} {904, 35}

\bibitem[\protect\citeauthoryear{{Perley}, {Taggart}, {Dahiwale}  \&
  {Fremling}}{{Perley} et~al.}{2021}]{2021TNSCR.341....1P}
{Perley} D.~A.,  {Taggart} K.,  {Dahiwale} A.,   {Fremling} C.,  2021,
  Transient Name Server Classification Report, \href
  {https://ui.adsabs.harvard.edu/abs/2021TNSCR.341....1P} {2021-341, 1}

\bibitem[\protect\citeauthoryear{{Perlmutter} et~al.,}{{Perlmutter}
  et~al.}{1999}]{1999ApJ...517..565P}
{Perlmutter} S.,  et~al., 1999, \mn@doi [\apj] {10.1086/307221}, \href
  {https://ui.adsabs.harvard.edu/abs/1999ApJ...517..565P} {517, 565}

\bibitem[\protect\citeauthoryear{{Phillips}}{{Phillips}}{1993}]{1993ApJ...413L.105P}
{Phillips} M.~M.,  1993, \mn@doi [\apjl] {10.1086/186970}, \href
  {https://ui.adsabs.harvard.edu/abs/1993ApJ...413L.105P} {413, L105}

\bibitem[\protect\citeauthoryear{{Piascik}, {Steele}, {Bates}, {Mottram},
  {Smith}, {Barnsley}  \& {Bolton}}{{Piascik}
  et~al.}{2014}]{2014SPIE.9147E..8HP}
{Piascik} A.~S.,  {Steele} I.~A.,  {Bates} S.~D.,  {Mottram} C.~J.,  {Smith}
  R.~J.,  {Barnsley} R.~M.,   {Bolton} B.,  2014, in {Ramsay} S.~K.,  {McLean}
  I.~S.,   {Takami} H.,  eds,  Society of Photo-Optical Instrumentation
  Engineers (SPIE) Conference Series Vol. 9147, Ground-based and Airborne
  Instrumentation for Astronomy V. p. 91478H, \mn@doi{10.1117/12.2055117}

\bibitem[\protect\citeauthoryear{{Planck Collaboration} et~al.,}{{Planck
  Collaboration} et~al.}{2020}]{2020A&A...641A...6P}
{Planck Collaboration} et~al., 2020, \mn@doi [\aap]
  {10.1051/0004-6361/201833910}, \href
  {https://ui.adsabs.harvard.edu/abs/2020A&A...641A...6P} {641, A6}

\bibitem[\protect\citeauthoryear{{Popovic}, {Brout}, {Kessler}, {Scolnic}  \&
  {Lu}}{{Popovic} et~al.}{2021}]{2021arXiv210201776P}
{Popovic} B.,  {Brout} D.,  {Kessler} R.,  {Scolnic} D.,   {Lu} L.,  2021,
  \mn@doi [\apj] {10.3847/1538-4357/abf14f}, \href
  {https://ui.adsabs.harvard.edu/abs/2021ApJ...913...49P} {913, 49}

\bibitem[\protect\citeauthoryear{{Riess} et~al.,}{{Riess}
  et~al.}{1998}]{1998AJ....116.1009R}
{Riess} A.~G.,  et~al., 1998, \mn@doi [\aj] {10.1086/300499}, \href
  {https://ui.adsabs.harvard.edu/abs/1998AJ....116.1009R} {116, 1009}

\bibitem[\protect\citeauthoryear{{Rigault}}{{Rigault}}{2018}]{2018zndo...1345222R}
{Rigault} M.,  2018, {ztfquery, a python tool to access ZTF data},
  \mn@doi{10.5281/zenodo.1345222}

\bibitem[\protect\citeauthoryear{{Rigault} et~al.,}{{Rigault}
  et~al.}{2015}]{2015ApJ...802...20R}
{Rigault} M.,  et~al., 2015, \mn@doi [\apj] {10.1088/0004-637X/802/1/20}, \href
  {https://ui.adsabs.harvard.edu/abs/2015ApJ...802...20R} {802, 20}

\bibitem[\protect\citeauthoryear{{Rigault} et~al.,}{{Rigault}
  et~al.}{2020}]{2020A&A...644A.176R}
{Rigault} M.,  et~al., 2020, \mn@doi [\aap] {10.1051/0004-6361/201730404},
  \href {https://ui.adsabs.harvard.edu/abs/2020A&A...644A.176R} {644, A176}

\bibitem[\protect\citeauthoryear{{Saunders} et~al.,}{{Saunders}
  et~al.}{2018}]{2018ApJ...869..167S}
{Saunders} C.,  et~al., 2018, \mn@doi [\apj] {10.3847/1538-4357/aaec7e}, \href
  {https://ui.adsabs.harvard.edu/abs/2018ApJ...869..167S} {869, 167}

\bibitem[\protect\citeauthoryear{{Schlafly} \& {Finkbeiner}}{{Schlafly} \&
  {Finkbeiner}}{2011}]{2011ApJ...737..103S}
{Schlafly} E.~F.,  {Finkbeiner} D.~P.,  2011, \mn@doi [\apj]
  {10.1088/0004-637X/737/2/103}, \href
  {https://ui.adsabs.harvard.edu/abs/2011ApJ...737..103S} {737, 103}

\bibitem[\protect\citeauthoryear{{Schlegel}, {Finkbeiner}  \&
  {Davis}}{{Schlegel} et~al.}{1998}]{1998ApJ...500..525S}
{Schlegel} D.~J.,  {Finkbeiner} D.~P.,   {Davis} M.,  1998, \mn@doi [\apj]
  {10.1086/305772}, \href
  {https://ui.adsabs.harvard.edu/abs/1998ApJ...500..525S} {500, 525}

\bibitem[\protect\citeauthoryear{{Schulze} et~al.,}{{Schulze}
  et~al.}{2020}]{Schulze2020a}
{Schulze} S.,  et~al., 2020, arXiv e-prints, \href
  {https://ui.adsabs.harvard.edu/abs/2020arXiv200805988S} {p. arXiv:2008.05988}

\bibitem[\protect\citeauthoryear{{Schwarz}}{{Schwarz}}{1978}]{1978AnSta...6..461S}
{Schwarz} G.,  1978, Annals of Statistics, \href
  {https://ui.adsabs.harvard.edu/abs/1978AnSta...6..461S} {6, 461}

\bibitem[\protect\citeauthoryear{{Scolnic} \& {Kessler}}{{Scolnic} \&
  {Kessler}}{2016}]{2016ApJ...822L..35S}
{Scolnic} D.,  {Kessler} R.,  2016, \mn@doi [\apjl]
  {10.3847/2041-8205/822/2/L35}, \href
  {https://ui.adsabs.harvard.edu/abs/2016ApJ...822L..35S} {822, L35}

\bibitem[\protect\citeauthoryear{{Scolnic} et~al.,}{{Scolnic}
  et~al.}{2018}]{2018ApJ...859..101S}
{Scolnic} D.~M.,  et~al., 2018, \mn@doi [\apj] {10.3847/1538-4357/aab9bb},
  \href {https://ui.adsabs.harvard.edu/abs/2018ApJ...859..101S} {859, 101}

\bibitem[\protect\citeauthoryear{{Scolnic} et~al.,}{{Scolnic}
  et~al.}{2020}]{2020ApJ...896L..13S}
{Scolnic} D.,  et~al., 2020, \mn@doi [\apjl] {10.3847/2041-8213/ab8735}, \href
  {https://ui.adsabs.harvard.edu/abs/2020ApJ...896L..13S} {896, L13}

\bibitem[\protect\citeauthoryear{{Skrutskie} et~al.,}{{Skrutskie}
  et~al.}{2006}]{Skrutskie2006a}
{Skrutskie} M.~F.,  et~al., 2006, \mn@doi [\aj] {10.1086/498708}, \href
  {https://ui.adsabs.harvard.edu/abs/2006AJ....131.1163S} {131, 1163}

\bibitem[\protect\citeauthoryear{{Soraisam}, {Matheson}  \& {Lee}}{{Soraisam}
  et~al.}{2021}]{2021arXiv210309937S}
{Soraisam} M.,  {Matheson} T.,   {Lee} C.-H.,  2021, \mn@doi [Research Notes of
  the American Astronomical Society] {10.3847/2515-5172/abf1f7}, \href
  {https://ui.adsabs.harvard.edu/abs/2021RNAAS...5...62S} {5, 62}

\bibitem[\protect\citeauthoryear{{Steele} et~al.,}{{Steele}
  et~al.}{2004}]{2004SPIE.5489..679S}
{Steele} I.~A.,  et~al., 2004, in {Oschmann} Jacobus~M. J.,  ed.,  Society of
  Photo-Optical Instrumentation Engineers (SPIE) Conference Series Vol. 5489,
  Ground-based Telescopes. pp 679--692, \mn@doi{10.1117/12.551456}

\bibitem[\protect\citeauthoryear{{Sullivan} et~al.,}{{Sullivan}
  et~al.}{2010}]{2010MNRAS.406..782S}
{Sullivan} M.,  et~al., 2010, \mn@doi [\mnras]
  {10.1111/j.1365-2966.2010.16731.x}, \href
  {https://ui.adsabs.harvard.edu/abs/2010MNRAS.406..782S} {406, 782}

\bibitem[\protect\citeauthoryear{{Suzuki} et~al.,}{{Suzuki}
  et~al.}{2012}]{2012ApJ...746...85S}
{Suzuki} N.,  et~al., 2012, \mn@doi [\apj] {10.1088/0004-637X/746/1/85}, \href
  {https://ui.adsabs.harvard.edu/abs/2012ApJ...746...85S} {746, 85}

\bibitem[\protect\citeauthoryear{{Thorp}, {Mandel}, {Jones}, {Ward}  \&
  {Narayan}}{{Thorp} et~al.}{2021}]{2021arXiv210205678T}
{Thorp} S.,  {Mandel} K.~S.,  {Jones} D.~O.,  {Ward} S.~M.,   {Narayan} G.,
  2021, arXiv e-prints, \href
  {https://ui.adsabs.harvard.edu/abs/2021arXiv210205678T} {p. arXiv:2102.05678}

\bibitem[\protect\citeauthoryear{{Tripp}}{{Tripp}}{1998}]{1998A&A...331..815T}
{Tripp} R.,  1998, \aap, \href
  {https://ui.adsabs.harvard.edu/abs/1998A&A...331..815T} {331, 815}

\bibitem[\protect\citeauthoryear{{Tripp} \& {Branch}}{{Tripp} \&
  {Branch}}{1999}]{1999ApJ...525..209T}
{Tripp} R.,  {Branch} D.,  1999, \mn@doi [\apj] {10.1086/307883}, \href
  {https://ui.adsabs.harvard.edu/abs/1999ApJ...525..209T} {525, 209}

\bibitem[\protect\citeauthoryear{{Wang}}{{Wang}}{2005}]{2005ApJ...635L..33W}
{Wang} L.,  2005, \mn@doi [\apjl] {10.1086/499053}, \href
  {https://ui.adsabs.harvard.edu/abs/2005ApJ...635L..33W} {635, L33}

\bibitem[\protect\citeauthoryear{{Wright} et~al.,}{{Wright}
  et~al.}{2010}]{Wright2010a}
{Wright} E.~L.,  et~al., 2010, \mn@doi [\aj] {10.1088/0004-6256/140/6/1868},
  \href {https://ui.adsabs.harvard.edu/abs/2010AJ....140.1868W} {140, 1868}

\bibitem[\protect\citeauthoryear{{Wright} et~al.,}{{Wright}
  et~al.}{2016}]{Wright2016a}
{Wright} A.~H.,  et~al., 2016, \mn@doi [\mnras] {10.1093/mnras/stw832}, \href
  {https://ui.adsabs.harvard.edu/abs/2016MNRAS.460..765W} {460, 765}

\bibitem[\protect\citeauthoryear{{van der Walt}, {Colbert}  \&
  {Varoquaux}}{{van der Walt} et~al.}{2011}]{2011CSE....13b..22V}
{van der Walt} S.,  {Colbert} S.~C.,   {Varoquaux} G.,  2011, \mn@doi
  [Computing in Science and Engineering] {10.1109/MCSE.2011.37}, \href
  {https://ui.adsabs.harvard.edu/abs/2011CSE....13b..22V} {13, 22}

\makeatother
\end{thebibliography}




\appendix
\section{Comparison to \texttt{SNooPy} fits}
\label{sec:app_snoopy}

\begin{figure}
    \centering
    \includegraphics[width=.5\textwidth]{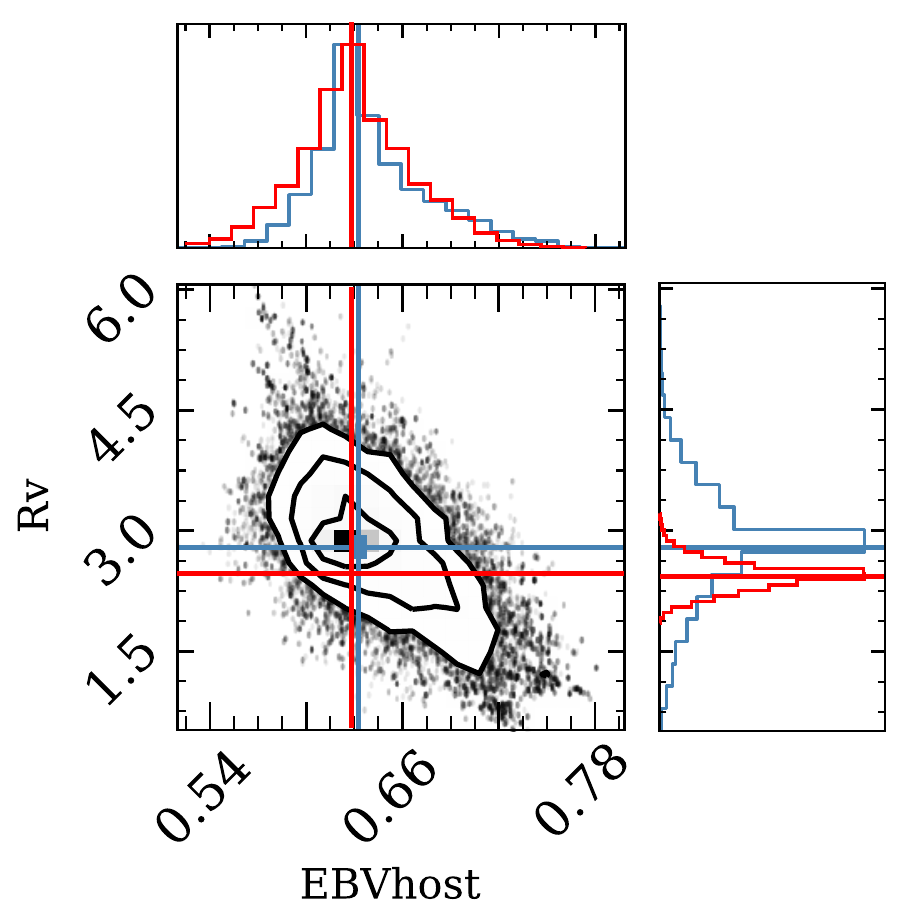}
    \caption{Contour plot showing the inferred $R_V$ and $E(B-V)$ for AT2019lcj using the SNooPy SN model without any prior (black), with a prior on the unextinguished peak magnitude from SN2020aewj (red). The best fit $R_V = 2.8 \pm 0.7$ and $E(B-V) = 0.63 \pm 0.07$ mag without any prior and $R_V = 2.45 \pm 0.18$ and an $E(B-V) = 0.63 \pm 0.07$ with a prior on the peak apparent magnitude from SN~2020aewj.}
    \label{fig:snpy_RvEBVcorner}
\end{figure}

For comparison with the SNe~Ia SALT2 colour model, we also infer the dust properties of the host galaxy using the \texttt{SNooPy} SN~Ia model. SNooPy is a modular software with different SN~Ia models, which can depend on various lightcurve shape parameters, to fit template lightcurves to SN~Ia data. While models like the \texttt{max model} only fit for the peak magnitude in each filter, we use the \texttt{color model}, which also fits for the reddening parameters $R_V$ and $E(B-V)$. The color model templates are a function of the colour-stretch, $s_{BV}$, which is the time since maximum when the $B-V$ colour curve reaches its maximum value, normalised to 30\,d. The $s_{BV}$ parameter is shown to be a better ordering parameter than the conventional $\Delta m_{15}$ lightcurve shape, especially for fast-declining SNe~Ia \citep{2014ApJ...789...32B}. 
Fitting the SNooPy colour model to the $g,r,i$ data for AT2019lcj yields an $R_V = 2.8 \pm 0.71$ and $E(B-V) = 0.63 \pm 0.07$ mag (see Figure~\ref{fig:snpy_RvEBVcorner}. Additionally, similar to the procedure with the SALT2 colour-luminosity and use a prior on the apparent reddening corrected magnitude of AT2019lcj from the model fit to SN2020aewj. This yields an $R_V = 2.45 \pm 0.18$ and an $E(B-V) = 0.63 \pm 0.07$ mag. The $R_V$ values from the SNooPy fit are consistent well within 1 $\sigma$ with the $\beta - 1$ from the SALT2 fits.
\pagebreak

\section{The Dust model in Brout \& Scolnic (2021)} 
\label{app:bs}

\begin{figure}
    \begin{center}
    \includegraphics[width=\columnwidth]{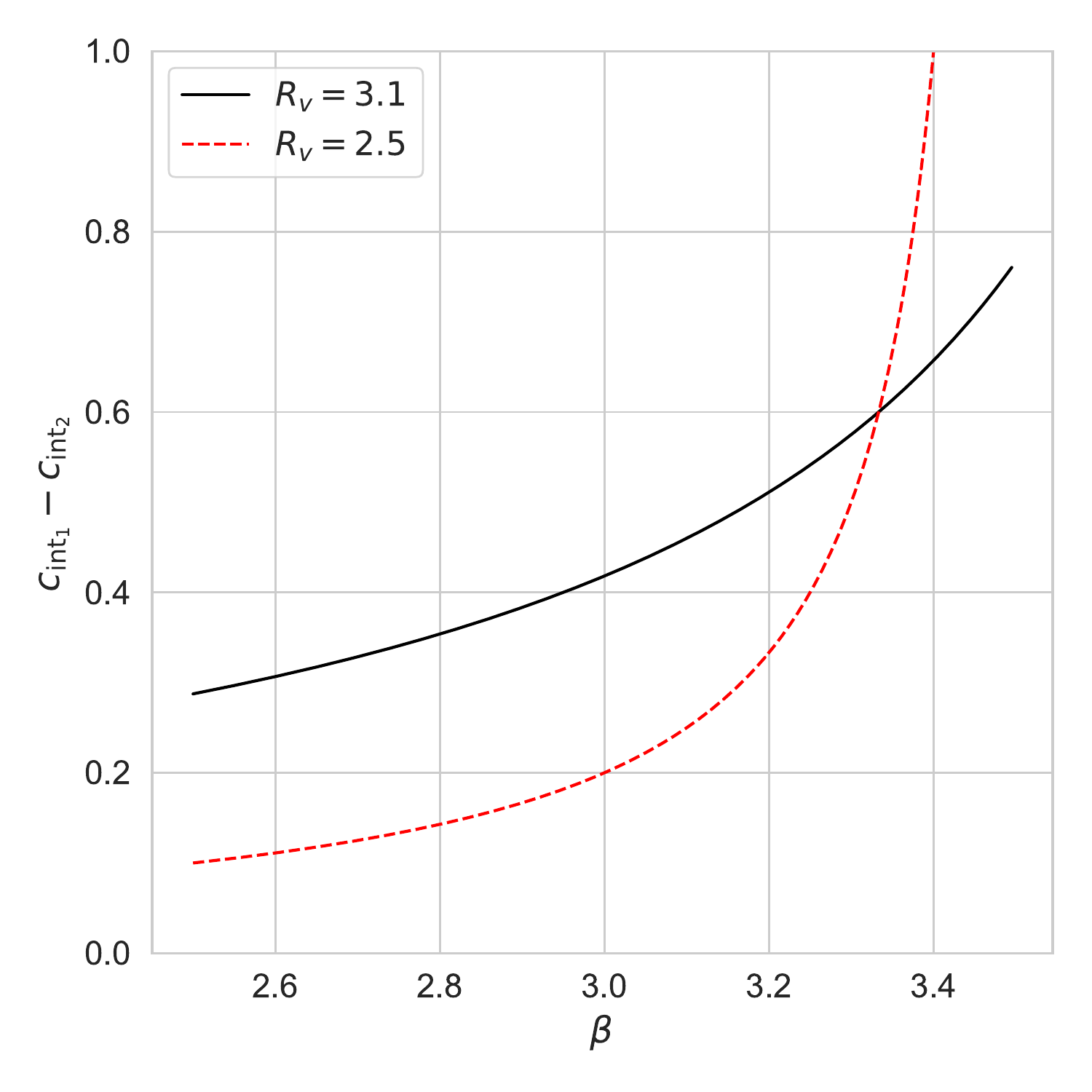}
    \caption{The differences in $c_{int}$ in the model described in ~\citet{2020arXiv200410206B} for the pair of siblings described as a function of $\beta$ (note that this is different from the \changes{SALT standarization model} $\beta$ constrained in Sec.~\ref{sec:results} for two choices of $R_V$.}
    \label{fig:cint}
    \end{center}
\end{figure}

~\citet{2020arXiv200410206B} propose a change to the prevalent \salt\ methodology positing that the \salt\ colour parameter $c$ obtained from light curve fits is the sum of two terms 
\be
c_{SALT}  = c_{int} + E_{dust}
\label{eqn:decomp}
\ee
where $E_{dust}$ is the redenning due to the host galaxy dust and is constrained to be positive, while $c_{int}$ is a colour intrinsic to the supernova. Accordingly, they modify the \changes{SALT standardization relation} 
\be 
\mu = m - M + \alpha \cdot x_1 - \beta \cdot c
\ee
to 
\be 
\mu = m - M + \alpha \cdot x_1 - \beta \cdot c_{int} - R_B \cdot E_{dust}
\label{eqn:bs}
\ee
demonstrating that a correlation between $R_B$ and stellar mass of the host galaxy can account for the observed host correlations. A side-effect is that the population distribution of $c_{int}$ is a fairly narrow Gaussian with a mean of $0.084$ and standard deviation of $0.042$. Thus, the well known red tail in supernova populations (see for example, ~\citet{ 2016ApJ...822L..35S}) is interpreted as being due to higher values of $E_{dust}$ in this model.\\

In the context of the current work, this model is different from the standard \salt\ model and \changes{SALT standardization relation}. It has more free parameters, and clearly has a different physical interpretation from the stanardard \salt\ and \changes{SALT standardization} relations.  In fact, here $\beta$ which has a universal value is completely independent of $R_B,$ which varies from host galaxy to host galaxy.~\citet{2020arXiv200410206B} demonstrate that a correlation between stellar masses of galaxies and $R_B$ coud drive the observed correlations between the standardized brightnesses of \snia\ and their host galaxies. We note that the stellar mass of the host galaxy of the siblings is comparable to that of the Milky Way, and thus one might expect this galaxy to have a similar $R_V$ as the Milky Way based on this model.\\

While we postpone a proper analysis of this model to future work, we can set the \salt\ parameters to the maximum likelihood of the \salt\ parameters as recorded in Tab.~\ref{tab:locations}.  
Approximating the individual fits in Tab.~\ref{tab:locations}, as the difference of \salt\ c parameters is $0.57,$  the difference in $x1 \sim 0,$ and the difference in $m_B^{\star}$  as $\sim 2$, we fix these parameeters. Using Eqn.~\ref{eqn:decomp} and Eqn.~\ref{eqn:dist_diff} to eliminate the terms involving $E_{dust}$, this imposes a relationship between the remaining variables $\alpha, \beta, R_B, c_{int}.$ Utilizing the Pantheon Priors on $\alpha,$ this gives us the difference in $c_{int},$ $\Delta c_{int} = c_{int}^1 - c_{int}^2$ for different values of $R_B$ and $\beta.$ 
In Fig.~\ref{fig:cint}, we show the values of $\Delta c_{int}$ for different choices $\beta$ where we have fixed $R_V$ to 3.1 or 2.5. Since we know the population distribution of $c_{int}$ is a normal distribution of known mean and variance, we can see that the pdf of the $\Delta c_{int}$ is a normal distribtion with mean approximately 1.6 and a standard deviation of 0.06.

\newpage
\section{Photometry Tables Used}
Note these tables are also made available in electronic format.
\begin{table*}
\caption{Light curve points for \oldsn\ used in the analysis. The zero points are in the AB system.}
\begin{tabular}{|c|c|c|c|c|c|c|}
jd (days) & band & flux (counts) & flux\_err (counts) & zp & $\mathrm{seeing} (\arcsec)$ & $m_\mathrm{lim}$ (mag) \\
\hline
2458655.79 & p48g & 18.72 & 37.23 & 25.93 & 2.9745 & 19.91 \\
2458662.94 & p48g & 56.02 & 22.66 & 26.06 & 1.9512 & 20.79 \\
2458663.97 & p48g & 65.35 & 35.57 & 25.58 & 2.1306 & 19.75 \\
2458663.97 & p48g & -58.13 & 36.12 & 25.54 & 2.174 & 19.64 \\
2458672.81 & p48g & 143.99 & 18.31 & 26.02 & 2.4073 & 20.73 \\
2458672.83 & p48g & 93.51 & 18.66 & 26.04 & 1.9574 & 20.96 \\
2458673.80 & p48g & 181.56 & 23.37 & 26.09 & 2.2867 & 20.62 \\
2458673.83 & p48g & 167.44 & 19.76 & 26.08 & 1.9766 & 20.94 \\
2458674.83 & p48g & 216.91 & 28.24 & 26.11 & 1.779 & 20.64 \\
2458674.85 & p48g & 266.67 & 22.44 & 26.08 & 2.0647 & 20.74 \\
2458675.81 & p48g & 183.23 & 36.75 & 26.07 & 1.9471 & 20.28 \\
2458675.85 & p48g & 243.35 & 31.69 & 26.06 & 1.8856 & 20.45 \\
2458678.78 & p48g & 369.44 & 52.74 & 26.06 & 2.6459 & 19.58 \\
2458679.79 & p48g & 354.50 & 58.68 & 26.09 & 2.1964 & 19.67 \\
2458680.70 & p48g & 497.51 & 65.23 & 26.08 & 2.1413 & 19.58 \\
2458681.70 & p48g & 353.79 & 66.10 & 26.07 & 2.3575 & 19.46 \\
2458682.70 & p48g & 416.38 & 51.36 & 26.09 & 2.0582 & 19.86 \\
2458683.78 & p48g & 404.32 & 49.62 & 26.09 & 1.9435 & 19.96 \\
2458684.77 & p48g & 412.83 & 41.90 & 25.99 & 2.5365 & 19.86 \\
2458685.87 & p48g & 430.13 & 43.94 & 26.02 & 2.2696 & 19.90 \\
2458693.77 & p48g & 353.75 & 22.05 & 26.05 & 2.2308 & 20.69 \\
2458696.76 & p48g & 275.83 & 22.50 & 26.08 & 1.998 & 20.75 \\
2458696.81 & p48g & 278.77 & 22.62 & 26.06 & 1.9962 & 20.75 \\
2458697.83 & p48g & 250.44 & 23.89 & 26.04 & 1.865 & 20.72 \\
2458699.81 & p48g & 165.57 & 24.34 & 26.01 & 2.0476 & 20.63 \\
2458700.76 & p48g & 236.76 & 23.21 & 26.07 & 2.0132 & 20.70 \\
2458700.81 & p48g & 252.79 & 23.53 & 26.05 & 1.8405 & 20.75 \\
2458701.81 & p48g & 112.43 & 23.90 & 25.96 & 1.9193 & 20.62 \\
2458702.81 & p48g & 145.08 & 24.18 & 25.97 & 1.8191 & 20.65 \\
2458703.69 & p48g & 106.19 & 34.28 & 25.93 & 2.6593 & 19.99 \\
2458703.81 & p48g & 201.14 & 24.65 & 25.99 & 1.9368 & 20.61 \\
2458704.81 & p48g & 136.22 & 34.00 & 26.01 & 2.3005 & 20.13 \\
2458704.88 & p48g & 122.88 & 26.77 & 25.89 & 2.7116 & 20.19 \\
2458704.90 & p48g & 134.62 & 27.42 & 25.87 & 2.7151 & 20.12 \\
2458705.73 & p48g & 127.43 & 41.91 & 26.03 & 1.7947 & 20.16 \\
2458705.91 & p48g & 163.97 & 27.59 & 25.88 & 2.4236 & 20.22 \\
2458706.68 & p48g & 123.65 & 45.22 & 26.11 & 1.951 & 20.07 \\
2458706.73 & p48g & 46.97 & 46.01 & 26.03 & 1.928 & 20.04 \\
2458707.91 & p48g & 125.95 & 48.56 & 25.97 & 2.0606 & 19.81 \\
2458708.66 & p48g & 97.68 & 55.14 & 26.07 & 1.6609 & 19.93 \\
2458708.73 & p48g & 189.82 & 55.55 & 26.05 & 2.2432 & 19.69 \\
2458708.89 & p48g & -9.06 & 62.65 & 25.97 & 2.0441 & 19.52 \\
2458709.73 & p48g & 151.77 & 61.24 & 26.10 & 1.9602 & 19.70 \\
2458710.66 & p48g & 206.42 & 63.87 & 26.10 & 1.6057 & 19.83 \\
2458710.71 & p48g & 97.73 & 68.72 & 26.10 & 2.0118 & 19.56 \\
2458710.73 & p48g & 92.19 & 69.68 & 26.05 & 1.9891 & 19.54 \\
2458711.71 & p48g & 72.08 & 62.48 & 26.08 & 1.7558 & 19.78 \\
2458711.76 & p48g & 77.31 & 65.87 & 26.07 & 1.8454 & 19.67 \\
2458714.71 & p48g & 67.30 & 37.72 & 26.04 & 1.8598 & 20.27 \\
2458715.80 & p48g & 15.07 & 42.89 & 26.02 & 1.8539 & 20.10 \\
2458716.82 & p48g & 47.07 & 40.81 & 26.00 & 1.9475 & 20.10 \\
2458717.80 & p48g & 106.93 & 34.90 & 25.98 & 2.1284 & 20.18 \\
2458718.82 & p48g & 80.48 & 31.08 & 25.91 & 2.5508 & 20.10 \\
2458719.80 & p48g & 85.00 & 26.32 & 25.90 & 2.511 & 20.25 \\
2458720.82 & p48g & 45.66 & 27.68 & 25.90 & 1.8968 & 20.43 \\
2458722.80 & p48g & 67.01 & 26.90 & 25.96 & 1.7849 & 20.55 \\
2458725.68 & p48g & 59.00 & 24.40 & 25.94 & 2.5194 & 20.40 \\
2458725.70 & p48g & 117.88 & 24.53 & 26.02 & 2.1454 & 20.57 \\
2458725.80 & p48g & 66.25 & 26.46 & 26.01 & 1.895 & 20.58 \\
2458726.80 & p48g & 8.70 & 27.05 & 25.94 & 2.2034 & 20.38 \\
2458727.77 & p48g & 40.43 & 26.43 & 25.97 & 1.8142 & 20.57 \\
2458727.82 & p48g & 12.92 & 28.40 & 25.93 & 1.9406 & 20.41 \\
2458728.68 & p48g & 57.67 & 25.33 & 25.96 & 1.6413 & 20.65 \\
2458730.80 & p48g & 16.70 & 28.23 & 25.85 & 2.0934 & 20.28 \\
\end{tabular}
\label{tab:oldsn}
\end{table*}
\begin{table*}
\contcaption{Light curve points for \oldsn\ used in the analysis}
\begin{tabular}{|c|c|c|c|c|c|c|}
\hline

2458732.65 & p48g & 48.68 & 34.33 & 25.98 & 1.8434 & 20.29 \\
2458732.68 & p48g & 19.28 & 34.25 & 25.98 & 1.6656 & 20.36 \\
2458732.69 & p48g & 8.11 & 33.99 & 25.97 & 1.8042 & 20.32 \\
2458732.79 & p48g & 29.79 & 26.97 & 25.95 & 1.7882 & 20.54 \\
2458733.66 & p48g & 11.55 & 36.81 & 26.03 & 1.9404 & 20.25 \\
2458733.79 & p48g & 96.79 & 33.92 & 25.89 & 2.5993 & 19.92 \\
2458733.81 & p48g & 61.58 & 31.29 & 25.91 & 2.3659 & 20.11 \\
2458733.83 & p48g & 45.68 & 29.15 & 25.84 & 2.6269 & 20.05 \\
2458734.69 & p48g & 91.92 & 39.74 & 25.96 & 2.7377 & 19.85 \\
2458734.72 & p48g & 13.64 & 40.51 & 25.92 & 2.9513 & 19.74 \\
2458734.80 & p48g & 33.17 & 40.78 & 25.95 & 2.3262 & 19.85 \\
2458735.79 & p48g & 59.31 & 47.94 & 25.93 & 2.1473 & 19.74 \\
2458736.80 & p48g & 3.06 & 55.10 & 25.85 & 2.8398 & 19.33 \\
2458659.68 & p48i & -98.84 & 37.22 & 25.52 & 1.3288 & 19.98 \\
2458663.68 & p48i & -39.71 & 28.13 & 25.45 & 1.8868 & 19.90 \\
2458672.94 & p48i & 131.23 & 30.23 & 25.48 & 1.8611 & 19.90 \\
2458679.76 & p48i & 438.90 & 36.02 & 25.56 & 1.5662 & 19.96 \\
2458685.80 & p48i & 293.17 & 35.28 & 25.52 & 1.461 & 19.98 \\
2458690.83 & p48i & 391.43 & 31.57 & 25.48 & 1.4088 & 20.12 \\
2458696.68 & p48i & 230.51 & 31.10 & 25.52 & 1.4778 & 20.15 \\
2458701.91 & p48i & 180.12 & 35.62 & 25.41 & 1.4528 & 19.91 \\
2458710.75 & p48i & 166.77 & 44.15 & 25.54 & 1.5235 & 19.73 \\
2458715.81 & p48i & 237.98 & 35.97 & 25.51 & 1.6284 & 19.90 \\
2458719.66 & p48i & 216.90 & 33.67 & 25.46 & 1.6936 & 19.86 \\
2458732.83 & p48i & 144.40 & 33.99 & 25.42 & 1.6258 & 19.83 \\
2458654.86 & p48r & 76.85 & 44.48 & 26.02 & 2.2151 & 19.92 \\
2458657.87 & p48r & -43.84 & 33.28 & 26.04 & 1.9803 & 20.39 \\
2458661.83 & p48r & -98.84 & 25.44 & 26.05 & 1.9287 & 20.69 \\
2458661.83 & p48r & 9.59 & 24.91 & 26.05 & 2.0323 & 20.70 \\
2458662.75 & p48r & 13.14 & 28.98 & 26.08 & 1.7381 & 20.59 \\
2458663.83 & p48r & -80.74 & 28.56 & 26.05 & 2.013 & 20.54 \\
2458663.84 & p48r & -6.23 & 28.78 & 26.06 & 1.966 & 20.56 \\
2458663.84 & p48r & 30.49 & 29.50 & 26.09 & 1.6037 & 20.67 \\
2458665.85 & p48r & 24.15 & 28.52 & 26.11 & 1.5575 & 20.70 \\
2458672.72 & p48r & 207.37 & 29.31 & 26.05 & 1.9369 & 20.51 \\
2458673.72 & p48r & 358.72 & 31.29 & 26.13 & 1.4066 & 20.64 \\
2458677.77 & p48r & 550.80 & 45.17 & 26.09 & 1.5020 & 20.26 \\
2458678.72 & p48r & 681.85 & 47.77 & 26.12 & 1.4399 & 20.24 \\
2458679.74 & p48r & 773.15 & 50.47 & 26.12 & 1.5577 & 20.15 \\
2458680.76 & p48r & 734.08 & 54.11 & 26.08 & 2.1612 & 19.77 \\
2458681.76 & p48r & 753.90 & 55.16 & 26.06 & 2.2524 & 19.73 \\
2458682.76 & p48r & 683.84 & 49.04 & 26.12 & 1.489 & 20.18 \\
2458683.75 & p48r & 782.34 & 44.65 & 26.10 & 1.5661 & 20.27 \\
2458684.74 & p48r & 697.07 & 37.29 & 26.02 & 2.2422 & 20.15 \\
2458685.84 & p48r & 830.17 & 40.62 & 26.09 & 1.5865 & 20.32 \\
2458688.89 & p48r & 864.20 & 38.46 & 26.03 & 1.523 & 20.34 \\
2458690.85 & p48r & 677.57 & 32.01 & 26.02 & 1.3362 & 20.56 \\
2458691.68 & p48r & 723.03 & 29.70 & 26.05 & 1.53 & 20.62 \\
2458692.83 & p48r & 594.66 & 32.16 & 26.04 & 1.6437 & 20.50 \\
2458693.72 & p48r & 574.09 & 31.92 & 26.03 & 1.5733 & 20.55 \\
2458694.70 & p48r & 579.28 & 30.94 & 26.10 & 1.395 & 20.68 \\
2458695.78 & p48r & 355.74 & 38.66 & 25.21 & 1.9794 & 19.30 \\
2458696.78 & p48r & 609.70 & 28.70 & 26.09 & 1.5222 & 20.70 \\
2458696.86 & p48r & 468.21 & 29.53 & 26.06 & 1.6503 & 20.60 \\
2458697.80 & p48r & 529.99 & 29.90 & 26.06 & 1.7317 & 20.59 \\
2458699.78 & p48r & 456.27 & 31.49 & 26.05 & 1.5015 & 20.56 \\
\end{tabular}
\label{tab:oldsn}
\end{table*}
\begin{table*}
\contcaption{Light curve points for \oldsn\ used in the analysis}
\begin{tabular}{|c|c|c|c|c|c|c|}
2458701.78 & p48r & 345.30 & 35.18 & 25.78 & 1.7682 & 20.08 \\
2458702.78 & p48r & 313.64 & 33.10 & 25.96 & 1.3461 & 20.48 \\
2458703.68 & p48r & 344.44 & 36.48 & 26.00 & 1.9797 & 20.24 \\
2458703.78 & p48r & 464.35 & 34.31 & 26.03 & 1.7509 & 20.39 \\
2458704.78 & p48r & 322.64 & 37.13 & 26.03 & 2.2181 & 20.11 \\
2458705.68 & p48r & 416.77 & 41.75 & 26.06 & 1.7215 & 20.24 \\
2458705.70 & p48r & 595.86 & 42.53 & 26.07 & 1.4837 & 20.30 \\
2458706.70 & p48r & 510.31 & 43.66 & 26.10 & 1.4447 & 20.29 \\
2458706.84 & p48r & 415.71 & 46.48 & 26.05 & 1.5837 & 20.16 \\
2458707.70 & p48r & 371.49 & 46.63 & 26.09 & 1.4241 & 20.26 \\
2458709.70 & p48r & 426.38 & 53.09 & 26.09 & 1.7109 & 20.03 \\
2458710.69 & p48r & 354.46 & 57.26 & 26.10 & 1.4671 & 20.01 \\
2458711.69 & p48r & 334.41 & 52.62 & 26.11 & 1.5541 & 20.07 \\
2458714.69 & p48r & 441.02 & 33.92 & 26.07 & 1.5714 & 20.55 \\
2458715.77 & p48r & 219.52 & 40.72 & 26.08 & 1.486 & 20.34 \\
2458716.79 & p48r & 275.66 & 40.50 & 26.06 & 1.654 & 20.28 \\
2458717.77 & p48r & 335.43 & 35.88 & 26.05 & 1.6434 & 20.41 \\
2458718.79 & p48r & 185.96 & 33.32 & 25.99 & 2.1706 & 20.25 \\
2458719.77 & p48r & 115.37 & 33.86 & 25.91 & 2.6513 & 20.00 \\
2458720.79 & p48r & 262.36 & 34.92 & 25.99 & 1.5167 & 20.42 \\
2458722.77 & p48r & 340.10 & 33.25 & 26.03 & 1.4654 & 20.50 \\
2458723.80 & p48r & 95.53 & 32.70 & 26.01 & 1.6926 & 20.50 \\
2458725.77 & p48r & 245.73 & 32.49 & 26.06 & 1.3984 & 20.56 \\
2458726.77 & p48r & 235.97 & 34.85 & 26.04 & 1.5069 & 20.49 \\
2458727.80 & p48r & 186.60 & 35.37 & 26.02 & 1.4836 & 20.42 \\
2458730.64 & p48r & 107.54 & 38.77 & 25.96 & 1.6622 & 20.23 \\
2458730.77 & p48r & 152.84 & 37.58 & 25.93 & 1.8356 & 20.15 \\
2458732.77 & p48r & 107.84 & 36.56 & 26.04 & 1.3995 & 20.44 \\
2458733.74 & p48r & 199.48 & 39.83 & 26.07 & 1.7258 & 20.28 \\
2458733.77 & p48r & 11.14 & 41.14 & 26.04 & 1.7149 & 20.23 \\
2458734.77 & p48r & 116.41 & 43.88 & 26.02 & 1.8755 & 20.06 \\
2458734.82 & p48r & 130.66 & 44.73 & 25.94 & 2.3806 & 19.78 \\
2458736.78 & p48r & 108.34 & 52.57 & 25.99 & 1.9647 & 19.84 \\
2458737.79 & p48r & 94.51 & 57.59 & 25.98 & 2.1669 & 19.64 \\
\end{tabular}
\label{tab:oldsn}
\end{table*}

\begin{table*}
\caption{Light curve points for \newsn\ used in the analysis. The zero points are in the AB system.}
\begin{tabular}{ccccccc}
jd (days) & band & flux (counts) & flux\_err (counts) & zp & $\mathrm{seeing} (\arcsec)$ & $m_\mathrm{lim}$ (mag) \\
2458877.02 & p48g & 742.07 & 27.51 & 26.08 & 2.6403 & 20.24 \\
2458880.98 & p48g & 1473.84 & 28.76 & 25.99 & 2.7017 & 20.21 \\
2458882.02 & p48g & 2005.69 & 27.94 & 26.12 & 2.2046 & 20.49 \\
2458891.97 & p48g & 2594.26 & 60.44 & 26.08 & 2.5028 & 19.51 \\
2458893.99 & p48g & 2401.47 & 46.10 & 26.09 & 2.1244 & 19.95 \\
2458899.05 & p48g & 2023.54 & 26.14 & 26.18 & 1.9566 & 20.74 \\
2458900.04 & p48g & 1777.22 & 27.06 & 26.10 & 2.427 & 20.38 \\
2458901.04 & p48g & 773.74 & 33.82 & 25.33 & 2.0688 & 19.53 \\
2458904.03 & p48g & 1337.40 & 24.73 & 26.15 & 2.0789 & 20.73 \\
2458909.05 & p48g & 852.24 & 44.56 & 26.16 & 2.3004 & 19.96 \\
2458912.05 & p48g & 600.80 & 33.44 & 26.03 & 2.8533 & 20.06 \\
2458914.02 & p48g & 456.64 & 26.50 & 25.93 & 2.0779 & 20.46 \\
2458914.94 & p48g & 400.14 & 58.13 & 25.97 & 2.0866 & 19.56 \\
2458915.97 & p48g & 443.95 & 66.20 & 26.03 & 2.2789 & 19.43 \\
2458942.00 & p48g & 187.59 & 20.76 & 26.00 & 2.8643 & 20.53 \\
2458874.03 & p48r & 320.95 & 34.97 & 26.14 & 1.8541 & 20.46 \\
2458876.05 & p48r & 457.65 & 30.19 & 26.04 & 2.5855 & 20.24 \\
2458877.06 & p48r & 735.35 & 34.38 & 26.18 & 1.7229 & 20.54 \\
2458878.02 & p48r & 887.10 & 32.10 & 26.10 & 2.205 & 20.34 \\
2458879.07 & p48r & 947.57 & 45.33 & 25.98 & 2.8889 & 19.71 \\
2458881.04 & p48r & 1502.91 & 34.15 & 26.11 & 2.2954 & 20.29 \\
2458882.05 & p48r & 1827.11 & 34.74 & 26.15 & 1.8957 & 20.46 \\
2458886.05 & p48r & 2295.86 & 36.98 & 26.14 & 2.0698 & 20.35 \\
2458887.06 & p48r & 2408.75 & 49.58 & 26.19 & 1.6498 & 20.22 \\
2458888.05 & p48r & 2480.49 & 63.14 & 26.18 & 1.7269 & 19.90 \\
2458894.05 & p48r & 2516.26 & 45.71 & 26.17 & 1.5729 & 20.34 \\
2458895.04 & p48r & 2546.97 & 42.90 & 26.18 & 1.6562 & 20.37 \\
2458895.99 & p48r & 2402.58 & 39.21 & 26.16 & 1.8695 & 20.35 \\
2458899.03 & p48r & 2148.01 & 32.88 & 26.17 & 1.5319 & 20.67 \\
2458899.98 & p48r & 1840.99 & 36.65 & 26.10 & 2.4675 & 20.10 \\
2458904.01 & p48r & 1585.96 & 32.71 & 26.16 & 1.8345 & 20.59 \\
2458908.99 & p48r & 1382.19 & 34.75 & 26.14 & 1.9151 & 20.46 \\
2458912.99 & p48r & 1361.77 & 38.80 & 26.09 & 1.7249 & 20.39 \\
2458914.00 & p48r & 975.57 & 39.22 & 25.84 & 1.5394 & 20.18 \\
2458937.02 & p48r & 598.16 & 36.08 & 26.13 & 1.784 & 20.48 \\
2458941.01 & p48r & 288.69 & 38.65 & 25.73 & 1.8366 & 19.96 \\
\end{tabular}
\label{tab:newsn}
\end{table*}



\bsp	
\label{lastpage}
\end{document}